\documentclass[12pt]{article}
\usepackage[english]{babel}
\usepackage{amssymb}
\usepackage{graphicx}
\usepackage{epsfig}

\def\Gtilde{{\widetilde\Gamma}}
\def\Qtilde{{\widetilde Q}}
\def\Ltilde{{\widetilde L}_0}
\def\Lhat{{\widehat L}_0}
\def\Dtilde{{\widetilde D}}
\def\Dcaltilde{\widetilde{\cal D}}
\def\Dhat{{\widehat D}}
\def\Mhat{{\widehat M}}
\def\Zhat{{\widehat Z}}
\def\be{\begin{equation}}
\def\ee{\end{equation}}
\def\bea{\begin{eqnarray}}
\def\eea{\end{eqnarray}}
\def\sss{\scriptscriptstyle}

\def\brs{\delta_{\rm BRS}}
\def\ss#1{{\sss (#1)}}
\def\Vtilde#1{{\widetilde V}_{\sss #1}}
\def\F#1{{F}_{\sss #1}}
\def\V#1{V_{\sss #1}}
\def\Wtilde#1{{\widetilde W}_{\sss #1}}
\def\Wcheck#1{\check{W}_{\sss #1}}
\def\W#1{W_{\sss #1}}
\def\h#1{{\mathrm h}^\ss{#1}}
\def\htilde#1{{\tilde{\mathrm{h}}}^\ss{#1}}
\def\H#1{{\cal H}^\ss{#1}}
\def\hcheck#1{{\check{\mathrm{h}}}^\ss{#1}}
\def\hhat#1{{\hat{\mathrm{h}}}^\ss{#1}}
\def\n#1{{\tilde\mathrm n}^{\sss (#1)}_{\sss -}}
\def\ncheck#1{{\check\mathrm{n}}^{\sss (#1)}_{\sss -}}
\def\O#1{\Omega_{\sss #1}}
\def\Q#1{{\rm\Qtilde}^{\sss (#1)}}
\def\Qbar#1{\overline{\rm\Qtilde}^{\vphantom{A}\atop{\vphantom{a}
\atop\!\!\sss (#1)}}\!}
\def\K#1{K^{\sss (#1)}}
\def\Zcheck{\check{Z}}

\begin{document}

\begin{titlepage}
\rightline{GEF-TH-9-2003 }
\vskip 1in
\begin{center}
\def\thefootnote{\fnsymbol{footnote}}
{\large Physical States at the Tachyonic Vacuum of Open\\ String
Field Theory}
\vskip 0.3in
S. Giusto\footnote{E-Mail: giusto@ge.infn.it} 
and C. Imbimbo\footnote{E-Mail: imbimbo@ge.infn.it} 
\vskip .2in
{\em Dipartimento di Fisica, Universit\`a di Genova\\
and\\ Istituto Nazionale di Fisica Nucleare, Sezione di Genova\\
via Dodecaneso 33, I-16146, Genoa, Italy}
\end{center}
\vskip .4in
\begin{abstract}
We illustrate a method for computing the number of physical states
of open string theory at the stable tachyonic vacuum in level truncation
approximation. The method is based on the analysis of 
the gauge-fixed open string field theory quadratic action that includes
Fadeev-Popov ghost string fields. Computations up to level 9 in
the scalar sector are consistent with Sen's conjecture about the
absence of physical open string states at the tachyonic vacuum.
We also derive a long exact
cohomology sequence that relates relative and absolute cohomologies
of the BRS operator at the non-perturbative vacuum. We use this
exact result in conjunction with our numerical findings to conclude
that the higher ghost number non-perturbative BRS cohomologies are non-empty.
\end{abstract}
\vfill
\end{titlepage}
\setcounter{footnote}{0}

\section{Introduction}

With the advent of D-branes it has been understood that bosonic open
strings are excitations of an unstable solitonic object of bosonic closed
string theory. This lead Sen to conjecture \cite{sen1,sen2} that the
non-linear classical equations of motion of open string field theory
(OSFT) \cite{witten} possess a translation invariant solution whose
energy density exactly cancels the brane tension.
The existence of a solution with such a property has been persuasively
demonstrated \cite{sz,mt,gr} within the level truncation (LT)
expansion of OSFT \cite{ks}. This solution, where the tachyon open
string field is condensed, is believed to be the stable
non-perturbative vacuum of OSFT representing the closed string vacuum
with no open strings.

The most basic expected property of OSFT around the tachyonic vacuum
is the absence of solutions of the linearized
equation of motions that are not pure gauge: 
this is what the conjecture that the stable
vacuum has no open string excitations means. The kinetic operator of
the OSFT action expanded around the non-perturbative vacuum solution
is a nilpotent operator $\Qtilde$ that acts on the first quantized
open string state space: the linearized equations of motion around the
tachyonic vacuum write in momentum space as
\be
\Qtilde(p)\,\psi^{(0)}(p) =0
\label{leom}
\ee
where $\psi^{(0)}(p)$ is an open string state of ghost number 0 and
space-time momentum $p$\footnote{We are adopting the convention in which
the $SL(2,\mathbb{R})$ invariant vacuum $|0\rangle$ has ghost number -1.}.  The space of solutions of the linearized
equations of motion (\ref{leom}) modulo (linearized) gauge
transformations is the cohomology of $\Qtilde(p)$ on open string
states of ghost number 0: it describes physical particles with mass
squared $m^2=-p^2$. Thus, in short, Sen's expectation is that the
cohomology  of $\Qtilde$ at ghost number 0 --- that we will denote by
$\H0(\Qtilde)$ --- vanishes for all $p^2$.

Computing the cohomology $\H0(\Qtilde)$ within the LT approximation
scheme faces one basic difficulty: by restricting the state space to states
of maximal level $L$ one breaks gauge invariance and
replaces $\Qtilde$ by a level truncated operator $\Qtilde_L$ which is
not nilpotent. Thus the image of $\Qtilde_L$ does not lie in the
kernel of $\Qtilde_L$. The task therefore is to understand which
solution of the level truncated linearized equations of motion should
be considered gauge-trivial.  The authors of \cite{et} proposed
to measure the triviality of a $\Qtilde_L$-closed state by its
orthogonal projection onto the image of $\Qtilde_L$.  This definition
requires a positive definite hermitian product, with respect to which the
orthogonal projection is defined. Although the open string state space
is equipped with a unique $\Qtilde$-invariant hermitian product, this is
not positive definite --- precisely because $\Qtilde$ is nilpotent.
For this reason the authors of \cite{et} select an {\it
arbitrary} positive definite hermitian product with respect to which
``approximate'' gauge-triviality of $\Qtilde_L$-closed states is
defined.

This way to compute $\H0(\Qtilde)$ may be subject to two kinds of
criticism. First, the choice of a positive definite 
hermitian product is arbitrary and it is not clear, {\it a priori},
why it should be irrelevant. The authors of \cite{et}
did verify that by taking two different positive definite hermitian
products their definition of ``approximate'' gauge-triviality
does not change much, but there is no theoretical reason why
this should be so in general: if a vector does not belong in
a given subspace of a vector space, it is always possible to
define a definite positive hermitian product with respect to which the vector 
and the subspace are orthogonal.

Second, any method that computes $\Qtilde$ cohomology by looking
at the kernel of the level truncated $\Qtilde_L$ can only
detect a certain type of cohomology. To understand this, let us
remark that the linear equations (\ref{leom}) have two kinds
of solutions: the $\psi^{(0)}(p)$'s that are solutions for $p$
{\it generic} and those that are solutions for isolated values
of $p$. We will call the former solutions of type A and the latter
solutions of type B. As we will review in Section 5, type B solutions are 
not gauge-trivial, while
type A solutions are gauge-trivial for $p$ generic, but they may become 
non-trivial at isolated values of $p$. Type B solutions are stable
in the LT approximation: they can be deformed but they are not expected
to disappear if the level is sufficiently big. Type A solutions instead
are lifted by the LT approximation for generic values of $p$ since 
$\Qtilde_L$ is not nilpotent: 
generically, the values of $p^2$ at
which a type A solution survives in the LT theory do not coincide
with the values of $p^2$ at which the solution of the exact theory may be
non-trivial. In summary, if one only looks at the solutions of
the level truncated version of the linearized equations of
motion (\ref{leom}), one misses cohomology of type A, if this
happens to exist. 

We will explain in Section 5 that $\Qtilde$-cohomology of type A
at ghost number $n$ is in one-to-one correspondence with cohomology
of type B at ghost number $n-1$.  For example, the {\it perturbative} 
linearized equations of motion (which
are the same as Eqs. (\ref{leom}) with $\Qtilde$ replaced by the BRS
charge $Q$ of the 2-dimensional CFT background) only
have non-trivial solutions of type B,
since the perturbative  BRS cohomology at ghost number -1 is empty 
(at non-exceptional momenta): we do not know of any reason to assume that
this holds also for the non-perturbative $\Qtilde$. 

For these reasons, we develop in this paper a method to
compute $\H0(\Qtilde)$ in the LT approximation which starts
from the {\it gauge-fixed} OSFT action around the tachyonic
vacuum, that we present in Section 2. 
In Siegel gauge, the gauge-fixed kinetic operator  
for the matter string field is the operator
\be
\Ltilde\equiv\{\Qtilde,b_0\}
\label{g-fkinetic}
\ee
acting on states of ghost number 0. The gauge-fixed
OSFT action includes also an infinite series of second quantized ghost
fields, whose kinetic operators are again given by the operator in
(\ref{g-fkinetic}) acting on states of non-vanishing ghost number $n$. 
The gauge-fixed kinetic operators for both 
matter and ghost string fields acting on
states of momentum $p$ will be denoted by $\Ltilde^\ss{n}(p)$, with
$n=0,1,\ldots$. 
The assumption that Siegel gauge is a good gauge
for the theory around the non-perturbative vacuum \cite{gr} 
means that the kinetic operators 
$\Ltilde^\ss{n}(p)$ do not have zero modes for generic values of $p^2$. 
$\Ltilde^\ss{n}(p)$ may have zero modes at isolated values of $p^2=-m^2$ and 
these zeros are connected with physical states of mass $m$. In a theory 
without gauge invariance, if the determinant 
$\det \Ltilde^\ss{0}(p)$ has a zero of order
$d$ for $p^2=-m^2$ there are $d$ physical degrees of freedom of mass $m$. 
In a gauge theory, ghost fields contribute to the physical degrees of freedom 
counting: if $\det \Ltilde^\ss{n}(p)$ has a
zero of order $d_n$ for $p^2=-m^2$, the number of physical degrees of freedom 
of mass $m$ --- which in OSFT is the dimension of $\H0(\Qtilde)|_{p^2=-m^2}$ 
--- is given by the Fadeev-Popov index:
\be
I_{FP}(m) = d_0 - 2\, d_1 + 2\, d_2 +\cdots = \sum_{n=-\infty}^\infty 
(-1)^n\, d_n
\label{fpindex0}
\ee

Our method to compute the dimension of $\H0(\Qtilde)$ in 
LT approximation consists in counting the zeros
of the determinants of the level truncated kinetic operators 
$(\Ltilde^\ss{n})_L (p)$. 
Of course degenerate zeros of $\det \Ltilde^\ss{n}(p)$
of the exact theory will correspond to zeros of the level truncated
$\det (\Ltilde^\ss{n})_L(p)$ that are only approximately degenerate. 
If the level is sufficiently big, we expect that zeros
of $\det (\Ltilde^\ss{n})_L(p)$ group into approximately degenerate
multiplets well separated among each other: 
in the level truncated theory, the index $I_{FP}$ in (\ref{fpindex}) should 
therefore be computed by 
including the zeros which belong to the same approximate multiplet.
Sen's conjecture is that Fadeev-Popov index for every multiplet
vanishes.
In order for this computation to make sense the level has
to be big enough that the splitting among zeros belonging to
the same multiplet is significantly smaller than the separation
between multiplets. 

In Section 3 we describe the results of our numerical study
of the operators $\Ltilde^\ss{n}$: we
restricted ourselves to the Lorentz scalar states to limit the numerical
complexity of the computation, although the extension of our method to
higher spin states is in principle straightforward and physically more
interesting.  We found that, for levels from 4 to 9, the LT evaluation
of the FP index $I_{FP}$ seems legitimate in the region $p^2\gtrsim
-5$: in this range of $p^2$ we found a single approximate multiplet at
$p^2 = -{\bar m}^2\approx -2.1$. The Fadeev-Popov index of this multiplet vanishes and we 
verified that its spreading decreases as the level goes up.

Since the zeros of the gauge-fixed kinetic operators
$\Ltilde^\ss{n}(p)$ of the exact theory are isolated, we expect them to
be more stable in LT than the zeros of the kinetic operator $\Qtilde
(p)$ of the gauge-invariant OSFT action. This is, in our view, the
main advantage of our method with respect to previous ones.  We expect
that in a fixed region of $p^2$, once the level is high enough, no new
zeros of the determinants $\det\Ltilde^\ss{n}(p)$ will flow in from
infinity as the level increases.  On the other hand, it might happen
that, for intermediate values of the level, {\it pairs} of zeros of the
level truncated $\det (\Ltilde^\ss{n})_L(p)$ disappear and reappear. We
found that for $p^2\approx -6$ there is indeed another multiplet of
zeros whose index vanishes for levels 5 and 6 and becomes 4 for
levels 7,8,9, due to the disappearance of a pair of zeros of 
$\det\Ltilde^\ss{-1}$: it is not inconceivable that at levels higher than 9
this pair of zeros reappears.

Our computation determined the quadratic part of the gauge-fixed 
OSFT action expanded
around the non-perturbative stable vacuum in the LT approximation. 
The knowledge of the kinetic operators $\Ltilde^\ss{n}(p)$, 
beyond allowing the test of Sen's conjecture that we just explained, 
provides further information about the dynamics of OSFT around the
non-perturbative vacuum. In this paper we exploited the knowledge
of the level truncated $\Ltilde^\ss{n}(p)$ to infer that the cohomology
of $\Qtilde$ does {\it not} vanish at ghost numbers -2 and -1. 

Our reasoning focused on the approximately degenerate multiplet of
zeros of $(\det\Ltilde^\ss{n})_L(p)$ located around $p^2 = -{\bar
m}^2\approx -2.1$.  Although the Fadeev-Popov index (\ref{fpindex0})
for this multiplet is zero, it turns out that the index constructed
with the {\it dimensions of the kernels} of $\Ltilde^\ss{n}(p)$ does
not vanish:
\be
\sum_{n=-\infty}^\infty (-1)^n \dim \ker\Ltilde^\ss{n}(p)|_{p^2 =-{\bar m}^2}
= 2
\label{relativeindex}
\ee
We will explain in Section 4 that this result provides information about the
dimensions of the ghost number $n$ cohomologies of $\Qtilde$ {\it relative} to the operator $b_0$. This relative cohomology, that we will denote
by $\htilde{n}$, is the cohomology of $\Qtilde$ evaluated on the subspace of 
$b_0$-invariant and $\Ltilde$-invariant states. 
The cohomologies of $\Qtilde$ on the
total space of ghost number $n$ are called instead {\it absolute}
cohomologies and they are denoted by $\H{n}(\Qtilde)$.
The ``experimental'' finding 
(\ref{relativeindex}) implies that
\be
\sum_{n=-\infty}^\infty (-1)^n \dim \htilde{n}|_{p^2 =-{\bar m}^2}
= 2
\label{relativeindexbis}
\ee
This means that, at $p^2 =-{\bar m}^2$, relative cohomologies
cannot vanish simultaneously for all ghost numbers $n$. 

In perturbative open string theory the knowledge of relative
cohomologies at all ghost numbers allows the computation of the
absolute BRS cohomologies $\H{n}(Q)$. Mathematically the relation
between relative and absolute perturbative BRS cohomologies is captured by a 
long exact sequence that we review in Section 4. In particular if the
relative cohomologies are non-vanishing at some ghost number the
absolute cohomologies cannot vanish at all ghost numbers. To reach a
similar conclusion in the non-perturbative case one has to understand
the relation between relative and absolute cohomologies of the
non-perturbative $\Qtilde$. We will do so in Section 4 where we will
derive a long exact sequence that relates the absolute
$\H{n}(\Qtilde)$'s, the relative $\htilde{n}$'s and another kind of
suitably defined relative cohomology. 
This is an exact result that is independent of the LT
approximation. We will show that taken together with  
(\ref{relativeindexbis}), the
non-perturbative long exact sequence implies 
\be 
\dim\H{-1}(\Qtilde)|_{p^2=-{\bar m}^2} =\dim \H{-2}(\Qtilde)|_{p^2=-{\bar
m}^2}=1 
\label{negativecohomology}
\ee 
in the Lorentz scalar sector of the theory. In Section 5 we provide
a consistency check of (\ref{negativecohomology}) which is independent from
the calculations of Section 4: it follows from the fact that the
cohomology of $\Qtilde(p)$ only appears on surfaces of positive 
codimension in momentum space.

Although our result
(\ref{negativecohomology}) does not contradict Sen's conjecture
it is in conflict both with the hypothesis of Vacuum SFT 
\cite{rsz} and with the numerical computations of \cite{efhm}.
It would be interesting to understand both the origin of this
discrepancy and the space-time significance of higher ghost number
cohomologies of $\Qtilde$.

\section{Gauge-fixed Open String Field Action}

The open string field theory (OSFT)  action 
around the tachyonic background writes
\be
\Gtilde[\Psi] = {1\over 2} \bigl( \Psi, \Qtilde\, \Psi\bigr) +
{1\over 3}\bigl(\Psi, \Psi\star\Psi\bigr)
\label{action}
\ee
$\Psi$ is the classical open string field, a state in the open
string Fock space of ghost number zero. $(A,B)$ is
the bilinear form between states $A$ and $B$ of ghost numbers $g_A$ and $g_B$
respectively. $(A,B)$ vanishes unless $g_A+g_B =1$.
$\star$ is Witten's
associative and non-commutative open string product.  
$\Qtilde$ is the BRS operator around the non-perturbative vacuum $\phi$
\be
\Qtilde\Psi \equiv Q\,\Psi + \bigl[\phi\,\mathop{,}^{\star} \Psi\bigr]
\ee
where
\be
\bigl[ A\, {\mathop{,}^{\star}} B\bigr]\equiv A\star B -
(-)^{(g_{\sss A}+1)\, (g_{\sss B}+1)}
B\star A
\ee
is the $\star$-(anti)commutator of string fields
$A$ and $B$ and
$Q$ is the perturbative BRS operator, which is (anti)symmetric with respect
to the bilinear inner product $(\,\cdot\,,\,\cdot\,)$ based on BPZ conjugation.
$\phi$ is the solution of the classical equation of motion
\be
Q\, \phi + \phi\star\phi =0
\label{flatness}
\ee
that represents the tachyonic vacuum. 
The flatness equation (\ref{flatness}),
together with the associativity of the $\star$-product, 
ensures the nilpotency of $\Qtilde$. $\Qtilde$ is (anti)symmetric with
respect to the product $(\,\cdot\,,\,\cdot\,)$ thanks to the property 
\be
(A,\,\phi\star B)=  (A\star\phi, B)
\ee
The action (\ref{action}) is
thus invariant under the following gauge transformations
\be
\delta\, \Psi = \Qtilde\, C + \bigl[\Psi\,\mathop{,}^{\star} C\bigr]
\label{gaugeinvariance}
\ee
where $C$ is a ghost number -1 gauge parameter.

Sen conjectured \cite{sen2} that the
translation invariant solution of Eq. (\ref{flatness}) $\phi$ 
represents the closed string vacuum with no D-branes: 
the classical open string field
action $\Gamma$ evaluated on such classical solution should equal
the tension of the D25 brane
\be
\Gamma[\phi] \equiv {1\over 2} \bigl( \phi, Q\, \phi\bigr) +
{1\over 3}\bigl(\phi, \phi\star\phi\bigr) = -{1\over 2\pi^2}
\label{tension}
\ee
The existence of a solution of the equation of motion (\ref{flatness})
satisfying (\ref{tension}) has been verified within the level expansion
scheme of OSFT with a good degree of accuracy 
\cite{sz,mt,gr}. 

The cohomology $\H0({\Qtilde})$ of the non-perturbative
BRS operator $\Qtilde$ on $\F0$, the space of states of ghost number 0, 
is to be
identified with the space of physical states around the tachyonic
vacuum. The physical interpretation of $\phi$ as the closed string
vacuum with no branes leads to the expectation \cite{sen1,sen2}
that this space is empty.

The fundamental difficulty in attempting to evaluate the cohomology
$\H0({\Qtilde})$ in the level truncated OSFT is that 
this approximation breaks
the gauge invariance of the action (\ref{action}): thus the problem
that one has to face is that of determining the (linearized) gauge-invariant
spectrum within a non-gauge invariant approximation scheme.

The LT approximation  consists in including a finite
number of open string states in the expansion
of the string field $\Psi$: if
\be
L_0 = {p^2\over 2} + {\hat {\rm N}} -1
\ee
is the Virasoro generator acting on open string states with space-time
momentum $p$ and ${\rm P}_L$ is the projector operator onto
the subspace with ${\hat{\rm N}}\le L$,
in theory truncated at level $L$ one replaces the string field 
$\Psi$ with its projection $\Psi_L \equiv {\rm P}_L \, \Psi$. 
Since 
\be
L_0 = \{Q,b_0\}
\ee
where $b_0$ is the zero mode of the antighost field $b(z)$, the level
commutes with the perturbative BRS operator $Q$. For this reason
the computation of the {\it perturbative} cohomology
can be restricted to the finite-dimensional
subspaces of given level --- this is, of course, what allows for 
the analytical solution of the problem. 
Unfortunately $L_0$ does not commute with the non-perturbative 
BRS operator
\be
\Qtilde_\phi \equiv Q + 
\bigl[\phi\,\mathop{,}^{\star}\cdot\,\bigr]
\ee
since  $L_0$ is not a derivative of the star product $\star$. 
Thus the projected operator 
\be
\Qtilde_L \equiv P_L\,\Qtilde_{\phi_{\sss L}}\, P_L
\ee
fails to be nilpotent 
\be
(\Qtilde_L)^2\not=0,
\ee
and the gauge invariance of the level truncated action is broken. 
Since $\Qtilde_L^2 \not=0$, the image of $\Qtilde_L$,  
is not contained in the kernel of $\Qtilde_L$, so that the concept
of cohomology of $\Qtilde_L$ does not make sense.
Our strategy will be to determine the physical spectrum of the OSFT
at the tachyonic vacuum by looking at the propagators obtained
by gauge-fixing the classical OSFT action. To this end, we will extend the
construction of \cite{bochicchiothorn} to the tachyonic theory.

CFT ghost number $g$ provides a grading for string fields: we have
seen that ``matter'' string field have $g=0$.  
We will also introduce another grading, the 
second quantized string field ghost number, that we will denote by 
$n_{\sss sft}$. Matter fields have $n_{\sss sft} =0$, by definition. 
Fields with second quantized ghost number $n_{\sss sft}=n$ and 
CFT ghost number $g$ will be denoted with $\Psi^{\ss{n}}_{\sss g}$.

The gauge invariance (\ref{gaugeinvariance}) of the classical OSFT action 
translates into the second quantized BRS symmetry
\be
\brs\, \Psi^{\ss0}_{\sss 0} = \Qtilde\, \Psi^{\ss{1}}_{\sss -1} + 
\bigl[\Psi^{\ss0}_{\sss 0}\,\mathop{,}^{\star} \Psi^{\ss1}_{\sss -1}\bigr]
\label{brsinvariance}
\ee
where $\Psi^{\ss1}_{\sss -1}$ is the ghost string field of first generation.
We will gauge-fix the invariance (\ref{brsinvariance}) by going to  
Siegel gauge:
\be
b_0 \, \Psi^{\ss0}_{\sss 0} = 0
\label{siegelgauge}
\ee
Thus the (partially) gauge-fixed Fadeev-Popov action
reads:
\be
\Gtilde_{g.f.} = {1\over 2} \bigl( \Psi^{\ss0}_{\sss 0}, \Qtilde\, \Psi^{\ss0}_{\sss 0}
\bigr)
+ {1\over 3}\bigl(\Psi^{\ss0}_{\sss 0}, \Psi^{\ss0}_{\sss 0}\star
\Psi^{\ss0}_{\sss 0}\bigr) +
\brs\,\bigl(B^{\ss{-1}}_{\sss 2}, b_0\, \Psi^{\ss0}_{\sss 0}\bigr)
\label{fpaction1}
\ee
where $B^{\ss{-1}}_{\sss 2}$ is the second quantized anti-ghost 
with $n_{sft}=-1$.
In order to compute propagators we will need the quadratic part of the
gauge-fixed action:
\be
\Gtilde_{g.f.}^{\ss2} = {1\over 2} \bigl( \Psi^{\ss0}_{\sss 0}, \Qtilde\, \Psi^{\ss0}_{\sss 0}
\bigr)+ 
\bigl(\Lambda^{\ss{0}}_{\sss 2}, b_0\, \Psi^{\ss0}_{\sss 0}\bigr) -
\bigl(B^{\ss{-1}}_{\sss 2}, b_0\,\Qtilde\, \Psi^{\ss{1}}_{\sss -1}\bigr)
\label{fpaction2}
\ee
where $\Lambda^{\ss{0}}_{\sss 2} =\brs\, B^{\ss{-1}}_{\sss 2}$ is the Lagrangian multiplier
enforcing the gauge choice (\ref{siegelgauge}). 

For any field $\Psi^\ss{n}_{\sss m}$ one can write the decomposition
\be 
\Psi^{\ss{n}}_{\sss m} = \phi^{\ss{n}}_{\sss m} + c_0\, \phi^\ss{n}_{\sss m-1}
\ee
where $\phi^\ss{n}_{\sss m}$ and $\phi^\ss{n}_{\sss m-1}$ 
are fields that do not contain $c_0$:
\be
b_0\, \phi^\ss{n}_{\sss m}\, = 0\qquad\forall\,\, m,\, n  
\ee
Integrating out 
$\Lambda^{\ss{0}}_{\sss 2}$ in the action (\ref{fpaction2}), one projects
 $\Psi^{\ss0}_{\sss 0}$ to its  $b_0$-invariant component $\phi^\ss0_{\sss 0}$.
Inserting $1 =\{c_0, b_0\}$
in the first term of Eq. (\ref{fpaction2}) and
introducing the tachyonic kinetic operator $\Ltilde$,
\be
\Ltilde\equiv \{ b_0 ,\Qtilde\}
\label{kinetictilde}
\ee
one obtains:
\bea
&&\Gtilde_{g.f.}^{\ss2} = {1\over 2} 
\bigl( \phi^{\ss0}_{\sss 0}, c_0\,b_0\, \Qtilde\, \phi^{\ss0}_{\sss 0}
\bigr)-
\bigl(B^\ss{-1}_{\sss 2}, b_0\,\Qtilde\, \Psi^\ss1_{\sss -1}\bigr)=\nonumber\\
&&\qquad = {1\over 2} 
\bigl(\phi^{\ss0}_{\sss 0}, c_0\, \Ltilde\, \phi^{\ss0}_{\sss 0}
\bigr)+
\bigl(\phi^\ss{-1}_{\sss 1},\Qtilde\, \Psi^\ss1_{\sss -1}\bigr)
\label{fpaction3}
\eea
where $\phi^\ss{-1}_{\sss 1} \equiv b_0 B^\ss{-1}_{\sss 2}$ 
does not contain $c_0$.

This action is still gauge-invariant under the (linearized) gauge
transformations
\be
\brs\,\Psi^\ss1_{\sss -1} = \Qtilde\, \Psi^\ss2_{\sss -2}
\ee
where $\Psi^{\ss2}_{\sss -2}$ is the second generation ghost string
field. To gauge-fix this
gauge-invariance we choose again the Siegel gauge 
\be
b_0 \Psi^\ss{1}_{\sss -1} = 0
\label{siegelgaugeminus1}
\ee
Introducing the anti-ghost
$B^\ss{-2}_{\sss 3}$ and integrating out the associated La\-gran\-gian multiplier 
$\Lambda^\ss{-1}_{\sss 3} =\brs\, B^\ss{-2}_{\sss 3}$, one obtains
\be
\Gtilde_{g.f.}^{\ss2} = {1\over 2} 
\bigl( \phi^{\ss0}_{\sss 0}, c_0\, \Ltilde\, \phi^{\ss0}_{\sss 0}
\bigr)+ 
\bigl(\phi^\ss{-1}_{\sss 1}, c_0\, \Ltilde\, \phi^{\ss1}\bigr)+
\bigl(\phi^\ss{-2}_{\sss 2},\Qtilde\, \Psi^{\ss2}_{\sss -2}\bigr)
\label{fpaction4}
\ee
where $\phi^\ss{-2}_{\sss 2}\equiv b_0\, B^\ss{-2}_{\sss 3}$. The action
(\ref{fpaction4}) requires further gauge-fixing: we will adopt the
Siegel gauge for all higher-generation ghost string fields:
\be
b_0\, \Psi^\ss{n}_{\sss -n} =0
\ee
Repeating the previous
steps one arrives to the following, completely gauge-fixed, quadratic action:
\bea
&& \Gtilde_{g.f.}^{\ss2} = {1\over 2} 
\bigl( \phi^{\ss0}_{\sss 0}, c_0\, \Ltilde\, \phi^{\ss0}_{\sss 0}
\bigr)+ \sum_{n=1}^\infty
\bigl(\phi^\ss{-n}_{\sss n}, c_0\, \Ltilde\, \phi^\ss{n}_{\sss -n}
\bigr)
\label{fpactioninfinity}
\eea 

Thus the gauge-fixed OSFT action depends on fields 
$\phi^\ss{-n}_{\sss n}\equiv \varphi_{\sss n}$ which are $b_0$-invariant
states of the first quantized Fock space with CFT ghost number $n$ and
second quantized ghost number $-n$. We will denote this state space 
with $\O{n}$.

It is convenient to define the following {\it non-degenerate} 
bilinear form $\langle \cdot,\cdot\rangle$ on $\O{-n} \times \O{n}$ 
\be
\langle \cdot, \cdot\rangle \equiv (\,\cdot\,, c_0\, \cdot\,)
\ee
From the definition  (\ref{kinetictilde}) of $\Ltilde$
and from the Jacobi identity one obtains:
\be
[\Ltilde, c_0] = [\{\Qtilde,b_0\},c_0] = [b_0, \{\Qtilde,c_0\}] = [b_0, \Dtilde]
\label{iacobi}
\ee
where $\Dtilde \equiv \{\Qtilde, c_0\}$. Note that the perturbative
kinetic operator $L_0$ commutes with $c_0$, and this is consistent
with the fact that the perturbative analogue of $\Dtilde$, $D= \{Q, c_0\}$ 
does not contain $c_0$. In the general non-perturbative case, the Jacobi identity 
only states that $\Ltilde$ and $c_0$ commute {\it up to a $b_0$-commutator}.
This is enough to ensure that $\Ltilde$ is an operator on
$\O{n}$ which is symmetric
with respect the bilinear form $\langle\, \cdot, \cdot\,\rangle$:
\be
\langle \varphi_{\sss n}, \Ltilde\, \varphi_{\sss -n}\rangle =
\langle\Ltilde\,  \varphi_{\sss n} , \varphi_{\sss -n}\rangle
\label{lhermitian}
\ee
In conclusion the quadratic part of the gauge-fixed OSFT action 
at the tachyonic background writes as
\be
\Gtilde_{g.f.}^{\ss2} =
{1\over 2} 
\langle \varphi_{\sss 0}, \Ltilde\, \varphi_{\sss 0}\rangle
+ \sum_{n=1}^\infty
\langle\varphi_{\sss n}, \Ltilde\, \varphi_{\sss -n}\rangle
\label{gaugefixedquadraticbis}
\ee
The field spaces $\O{n}$ can be decomposed as direct
sum of spaces with fixed space-time momentum $p^\mu$, $\mu=0,1,\ldots,25$:
\be
\O{n}  = \oplus_{p}\, \O{n}(p)
\ee
Because of translation invariance the kinetic operator $\Ltilde$
is diagonal with respect to this decomposition. For each space $\O{n}(p)$
choose a basis $\{e^{\sss (n)}_{i_n}(p)\}$. Let us denote
by $\Ltilde^\ss{n}(p)$ the matrix representing in this basis the operator
$\Ltilde$ acting on $\O{n}(p)$. 
Let $G^{\sss (n)}(p)$ be the square matrix whose elements are given by
\be
(G^{\sss (n)}(p))_{i_n\,j_n} = \langle e^{\sss (-n)}_{i_n}(p)\, ,\, 
e^{\sss (n)}_{j_n}(p)\rangle
\ee
For $n>0$ the symmetric square matrix that 
specifies the kinetic operator for the fields 
$(\varphi_{\sss -n},\varphi_{\sss n})$ is
\be
C^{\sss (-n)}(p) \equiv {1\over 2}
\left(\matrix{0& G^{\sss (n)}(p)\, \Ltilde^{\ss{n}}(p)\cr
G^{\sss (-n)}(p)\,\Ltilde^{\ss{-n}}(p) & 0\cr}\right)
\label{covariancen}
\ee
For the ``matter'' string field $\varphi_{\sss 0}$ the kinetic
quadratic form is instead
\be
C^{\sss (0)} \equiv  G^{\sss (0)}(p)\,\Ltilde^{\ss{0}}(p)
\label{covariancezero}
\ee

\section{Physical States via Fadeev-Popov Determinants}

The physical states of the OSFT at the tachyonic vacuum
can be read off the quadratic action (\ref{gaugefixedquadraticbis}).

Our analysis will focus on the determinants
\be
\Delta^\ss{n}(p^2) \equiv \det \Ltilde^\ss{n}(p)
\ee
which are functions of $p^2$. Were not for gauge-invariance, physical
states would correspond to zeros of $\Delta^\ss0(p^2)$: if 
$\Delta^\ss0(-m^2)=0$  and 
\be
\Delta^\ss0(p^2) = a_0\, (p^2+m^2)^{d_0} (1+ O(p^2+m^2))
\ee
there would be $d_0$ physical states with mass $m$. Ghosts fields
change the counting. Suppose that 
\be
\Delta^\ss{n}(p^2) = \Delta^\ss{-n}(p^2)  
= a_n\, (p^2+m^2)^{d_n} (1+ O(p^2+m^2))
\ee
where the first equality is a consequence of the symmetry property
(\ref{lhermitian}) of $\Ltilde$. 
Then, the number of physical states of mass $m$ is given by the
index:
\be
I_{FP}(m) = d_0 - 2\, d_1 + 2\, d_2 +\cdots = \sum_{n=-\infty}^\infty 
(-1)^n\, d_n
\label{fpindex}
\ee
This is so since the ghost and anti-ghost pairs $(\varphi_{\sss -n},
\varphi_{\sss n})$ are complex fields of Grassmanian parity  $(-1)^n$.
The numbers $d_n$ are gauge-dependent --- in our case they capture
properties of the $b_0$-invariant spaces $\O{n}$. The index $I_{FP}(m)$ 
is gauge-invariant and coincides with the dimension of the cohomology 
$\H0(\Qtilde)$ of $\Qtilde$ on $\F0$, the total  
space of (non-$b_0$-invariant) states of ghost number 0.
In a physical sensible theory $I_{FP}(m)$ must be non-negative. 
Sen's conjecture is that $I_{FP}$ vanishes for all $m$.

Let us see how the index formula in (\ref{fpindex}) works
in the {\it perturbative} theory.
In this case $L_0 = p^2/2 + {\hat {\rm N}}-1$ and thus
\be
\det L_0^\ss{n}(p) =0\qquad {\rm for}\quad p^2=-m_N^2 = 2(1-N)
\ee
with $N$ non-negative integer. For these values of $p^2$ the numbers $d_n$
are simply the dimensions of the subspaces of 
$\O{n}(p)$ with $p^2 = -m_N^2$ of level ${\hat {\rm N}} =N$. 
Let us denote $d_n$ at $p^2=-m_N^2$ with $d_n(N)$.
The generating function of the numbers $I_{FP}(m_N)$ of physical states
of mass $m_N$
\be
\sum_{N=0}^\infty I_{FP}(m_N)\,q^N
\ee 
equals, thanks to the Fadeev-Popov formula (\ref{fpindex}),
the function
\be
\sum_{N=0}^\infty\sum_{n=-\infty}^\infty (-1)^n\, d_n(N)\, q^N
\ee
The computation of this expression is standard. One can write
\be
d_n(N) = \sum_{k=0}^N d^{\sss (mat)}(k)\,d^{\sss (gh)}_n(N-k)
\ee
where $d^{\sss (mat)}(k)$ is the number of states of level $k$
in the bosonic Fock space
while $d^{\sss (gh)}_n(N-k)$ is the number of states in the
ghost Fock space with ghost number $n$ and level $N-k$. The generating
function for $d^{\sss (mat)}(k)$ is
\be
\sum_{k=0}^\infty d^{\sss (mat)}(k)\, q^k = \prod_{m=1}^\infty {1\over 
(1-q^m)^{26}}
\ee
and the generating function for the $d^{\sss (gh)}_n(N-k)$ is
\be
\sum_{k=0,\, n=0}^\infty d^{\sss (gh)}_n(k)\, q^k\, z^n = \prod_{m=1}^\infty
(1+ z\, q^m)( 1+{1\over z}\, q^m)
\ee
where in the R.H.S. of the equation above the two factors are the contributions
of the creator operators $c_{-n}$ and $b_{-n}$. Thus
\bea
&&\!\!\!\!\!\!\!\!\!\!\sum_{N=0}^\infty\sum_{n=-\infty}^\infty\!\!
(-1)^n\, d_n(N)\, q^N\!
=\!\sum_{N=0}^\infty\sum_{k=0}^N d^{\sss (mat)}(k)\, q^k\!\!\!
\sum_{n=-\infty}^\infty\!
(-1)^n\,d^{\sss (gh)}_n(N-k)\, q^{N-k}=\nonumber\\
&&\qquad\qquad\qquad\qquad\quad=\sum_{k=0}^\infty\, d^{\sss (mat)}(k)\, q^k\,
\sum_{N=0}^\infty\sum_{n=-\infty}^\infty
(-1)^n\,d^{\sss (gh)}_n(N)\, q^{N}=\nonumber\\
&&\qquad\qquad\qquad\qquad\quad=\prod_{m=1}^\infty {(1-q^m)^2\over(1-q^m)^{26}}=
\prod_{m=1}^\infty {1\over(1-q)^{24}}
\eea 
which is indeed the generating function for the physical states of 
bosonic open string theory in 26 dimensions.


Let us come back to the general non-perturbative case. 
Typically, in the exact (not level truncated) theory,
$\Delta^\ss{n}(p^2)$'s with different
ghost numbers $n$ vanish at the same value of $p^2$, as a consequence
of BRS invariance. Indeed $[\Qtilde, \Ltilde] = 0$; so, if
$\Delta^\ss{n}(p^2)$ vanishes for some $p^2=-m^2$, then
there exists a $\varphi_{\sss n}$ such that
\be
\Ltilde\,\varphi_{\sss n}= 0 = b_0 (\Qtilde\,\varphi_{\sss n})
\ee
Therefore 
\be
\Ltilde(\varphi_{\sss n+1})= 0 = b_0\,\varphi_{\sss n+1}
\ee
where  $\varphi_{n+1}=\Qtilde\,\varphi_{\sss n}$. If $\varphi_{n+1}$
does not vanish, $\Delta^\ss{n+1}(-m^2)=0$. Thus 
physical states of mass $m^2$ are associated to a multiplet of
determinants $\Delta^\ss{n}(p^2)$ with different $n$'s that vanish
simultaneously at $p^2=-m^2$. 
Since level truncation breaks BRS invariance we expect
that the zeros of the determinants in the same multiplet, 
when evaluated at finite $L$,
would be only approximately coincident. Thus using the index
formula (\ref{fpindex}) to compute the number of physical states
is meaningful when the splitting between approximately coincident
determinant zeros is significantly smaller than the distance between
the masses of different multiplets. 

\subsection{The numerical situation}

In the theory truncated at level $L$, the operators $\Ltilde^\ss{n}(p)$
reduce to finite dimensional matrices; moreover for a given $L$, the
$\Ltilde^\ss{n}(p)$ vanish identically for $n$ greater than a certain
$n_L$ which depends on the level\footnote{$n_L$ is the greatest integer which
satisfies the inequality  $n_L(n_L+1)/2 \le L$.}.
We  evaluated the LT matrices $\Ltilde^\ss{n}(p)$ on 
${\O{n}}^{\sss \!\!\!scalar}(p)$, the subspace of $\O{n}(p)$ containing
the states which are scalars with respect to space-time Lorentz symmetry
\footnote{The decomposition of $\O{n}(p)$ into irreducible
representations of the Lorentz group is legitimate only for $p^2\not=0$.
If one were interested in computing massless states one should in
principle include states of every spin. The number of physical
massless states is given by the sum of $I_{FP}(0)$ evaluated for each subspace
of states of given spin. A massless spin 1 state, for example, would
yield an index $I^{\sss vector}_{FP}(0)= 25$ in the vector sector
and $I^{\sss scalar}_{FP}(0)=-1$ in the scalar sector.}.
The extension of our methods to higher-spin states, in principle
straightforward, would be of considerable physical interest 
in particular to establish
the fate of the massless gauge boson.  It is however left for 
the future since, from the computational point of view, is relatively
demanding. 

The computation is simplified by noting that the non-perturbative
$\Qtilde$ commutes with the twist parity operator $(-1)^{\hat{\rm N}}$.
Therefore the kinetic operators decompose as follows
\be
\Ltilde^\ss{n}(p) = \Ltilde^\ss{n,+}(p)\oplus\Ltilde^\ss{n,-}(p)
\ee
where $\Ltilde^\ss{n,\pm}(p)$ are the kinetic operators acting
on the subspaces $\O{n}^\ss{\pm}(p)$ of $\O{n}(p)$
with twist parity $\pm$. 

Another symmetry of $\Ltilde$ is the $SU(1,1)$ symmetry generated
by:
\bea
&&J_+ = \{Q,c_0\} =\sum_{n=1}^\infty n\, c_{-n} c_n \qquad J_- =
\sum_{n=1}^\infty {1\over n}\, b_{-n} b_n\nonumber\\
&&J_3 = {1\over 2}\,
\sum_{n=1}^\infty (c_{-n} b_n - b_{-n} c_n)
\label{sutwo}
\eea
$J_\pm$ and $J_3$ are derivatives of the $\star$-product. They obviously
commute both with $b_0$ and the perturbative $L_0$ and hence they
are a symmetry of the OSFT equations of motion in the Siegel gauge:
\be
L_0\, \phi + b_0(\phi\star\phi)=0
\ee
The tachyon solution turns out to be a {\it singlet} of the $SU(1,1)$ algebra: 
it follows that  $J_\pm$ and $J_3$ commute with $\Ltilde$ since
\be
\Ltilde =L_0 + \{b_0,[\phi\,
\mathop{,}^{\!\!\!\star}\,\cdot]\}
\ee
Thus the multiplets of determinants $\Delta^\ss{n}(p^2)$ that vanish
at a given $p^2=-m^2$ organize themselves into representations 
of $SU(1,1)$. The symmetry (\ref{sutwo}) is not broken
by LT since its generators commute with the level:
therefore the $SU(1,1)$ symmetry of the multiplets of vanishing determinants
$\Delta^\ss{n}(-m^2)$ is exact even at finite $L$. 

We computed numerically the matrices $\Ltilde^\ss{n}(p)$ as functions
of $p$ on the subspaces ${\O{n}}^{\sss\!\!\!scalar}(p)$ in the theory
truncated at various levels $L$, from $L=4$ up to $L=9$.
Following \cite{sz},
we define the LT approximation to be of type $(L,M)$ if the 
OSFT action is restricted to fields up to level $L$ and
includes couplings between fields the sums of whose levels do not
exceed $M\leq 3 L$. For $L=4,\ldots,7$ our computation is of type $(L,3L)$: 
because of limitation of computational power at our disposal we performed
a computation of type $(L,2L)$ for the levels $L=8,9$. 
Thus for $L=4,\ldots,7$ we used the tachyon solution $\phi$ of level $(L,3L)$
while for $L=8,9$ we used a tachyon of level $(L,2L)$.

For $L\le 9$ the subspaces ${\O{n}}^{\sss\!\!\!scalar}(p)$
are non-empty for $|n|\le 3$. The dimensions of the matrices
$\Ltilde^\ss{n,+}(p)$ ($\Ltilde^\ss{n,-}(p)$) 
at even (odd) levels are listed in Table I. The dimension of the
matrix $\Ltilde^\ss{n,-}(p)$ ($\Ltilde^\ss{n,+}(p)$)
at the even (odd) level $L$ equals the
dimension of $\Ltilde^\ss{n,-}(p)$ ($\Ltilde^\ss{n,+}(p)$)
at level $L-1$. Since the tachyon $\phi$ only contains states
of even level, the matrices $\Ltilde^\ss{n,+}(p)$ of
level  $(4, 12)$ and $(6,12)$ are equal respectively to the matrices 
$\Ltilde^\ss{n,+}(p)$
of level $(5,15)$ and $(7,21)$. Other $\Ltilde^\ss{n,\pm}$ 
matrices with same dimensions differ among themselves 
since the tachyon couplings
that contribute to their non-perturbative parts are different.
$$
\vbox{\tabskip=0pt
\setbox\strutbox=\hbox{\vrule height12pt depth8pt width0pt}
\halign{\strut#& \vrule#& \hfil#\hfil &
\vrule#& \hfil#\hfil&\vrule#& \hfil#\hfil & \vrule#& \hfil#\hfil & \vrule#& \hfil#\hfil & \vrule#\tabskip=0pt\cr
\noalign{\hrule}
& & Level & & ~ghost \# 0~ & &~ghost \# -1~ & &~ghost \# -2~ & &~ghost \# -3
~&\cr\noalign{\hrule}
& & ~3 (odd)  & & 9   & & 6 & & 1 & & 0 &\cr\noalign{\hrule}
& & ~4 (even) & & 24  & & 13 & & 2 & & 0 &\cr\noalign{\hrule}
& & ~5 (odd) & & 45  & & 30 & & 7 & & 0 &\cr\noalign{\hrule}
& & ~6 (even) & & 99  & & 61 & & 14 & & 1 &\cr\noalign{\hrule}
& & ~7 (odd)  & & 183 & & 125 & & 35 & & 2 &\cr\noalign{\hrule}
& & ~8 (even) & & 363 & & 240 & & 68 & & 7 &\cr\noalign{\hrule}
& & ~9 (odd)  & & 655 & & 458 & & 145 & & 15 &\cr\noalign{\hrule}
}}
$$
\nobreak
\centerline{Table I: Number of $b_0$-invariant scalar states at various 
levels.}
\bigskip
  
\begin{figure}
\begin{center}
\includegraphics*[scale=.8, clip=false]{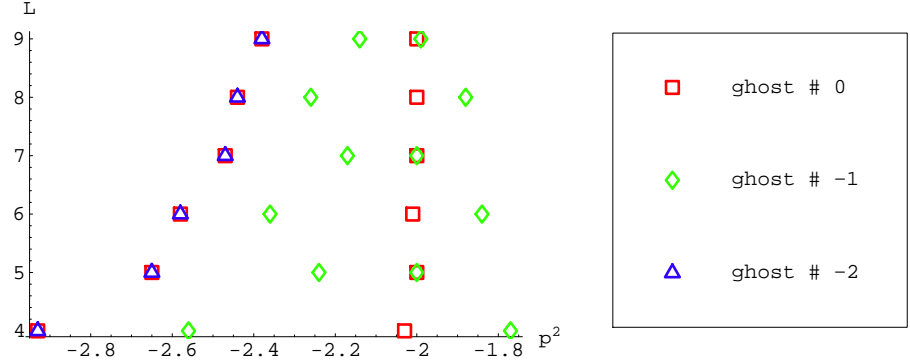}
\end{center}
\caption[x] {\footnotesize The first group of zeros of 
$\Delta^\ss{n}_{-}(p^2)$, for $n=0,-1,-2$ at levels $L= 4,\ldots,9$.}
\label{f:zeroes_gaugefixed.eps}
\end{figure}

\begin{figure}
\begin{center}
\raisebox{10pt}{$\scriptstyle{\rm (a)}$}
\includegraphics*[scale=.7, clip=false]{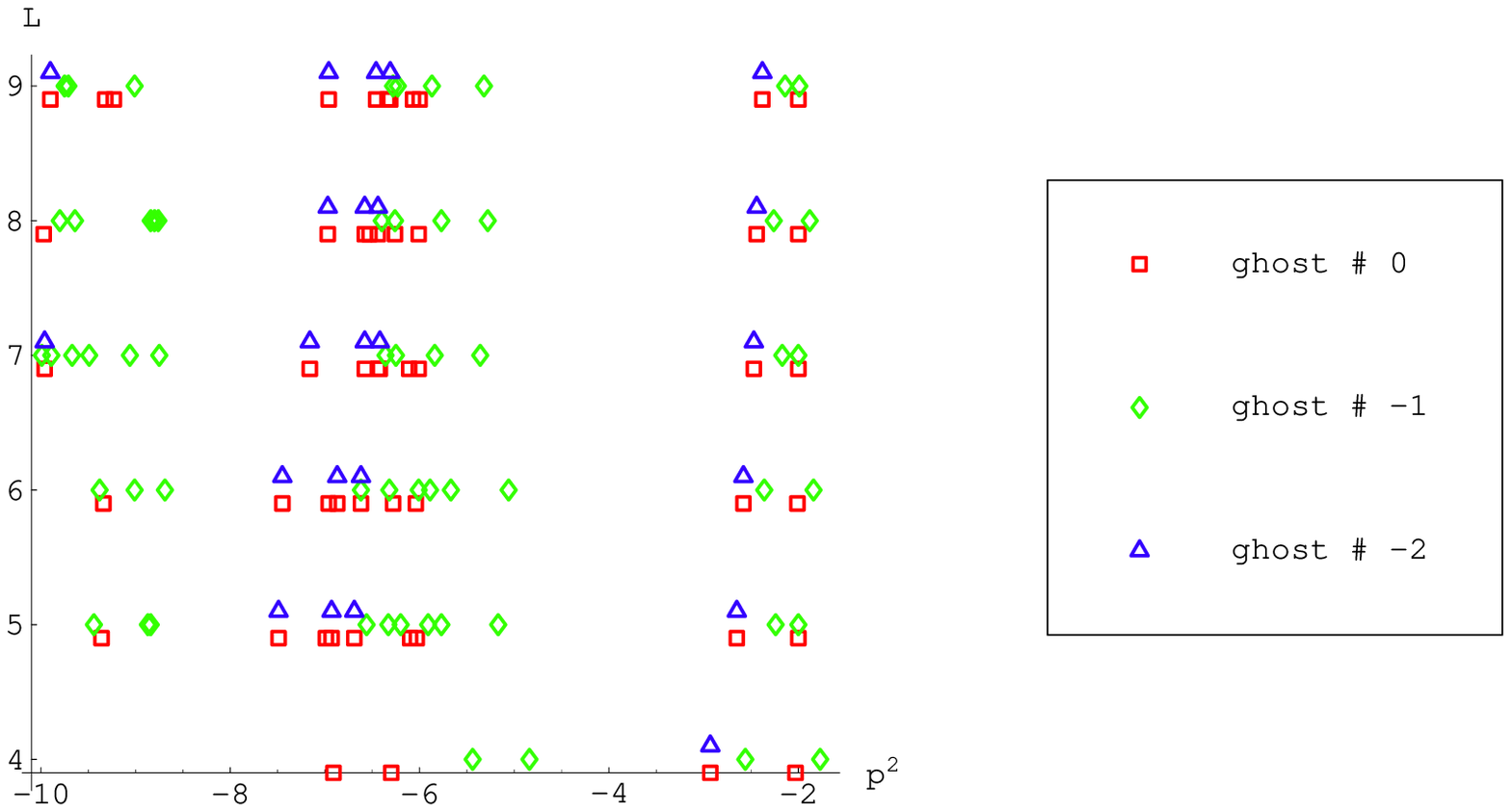}\\
\bigskip
\raisebox{10pt}{$\scriptstyle{\rm (b)}$}
\includegraphics*[scale=.7, clip=false]{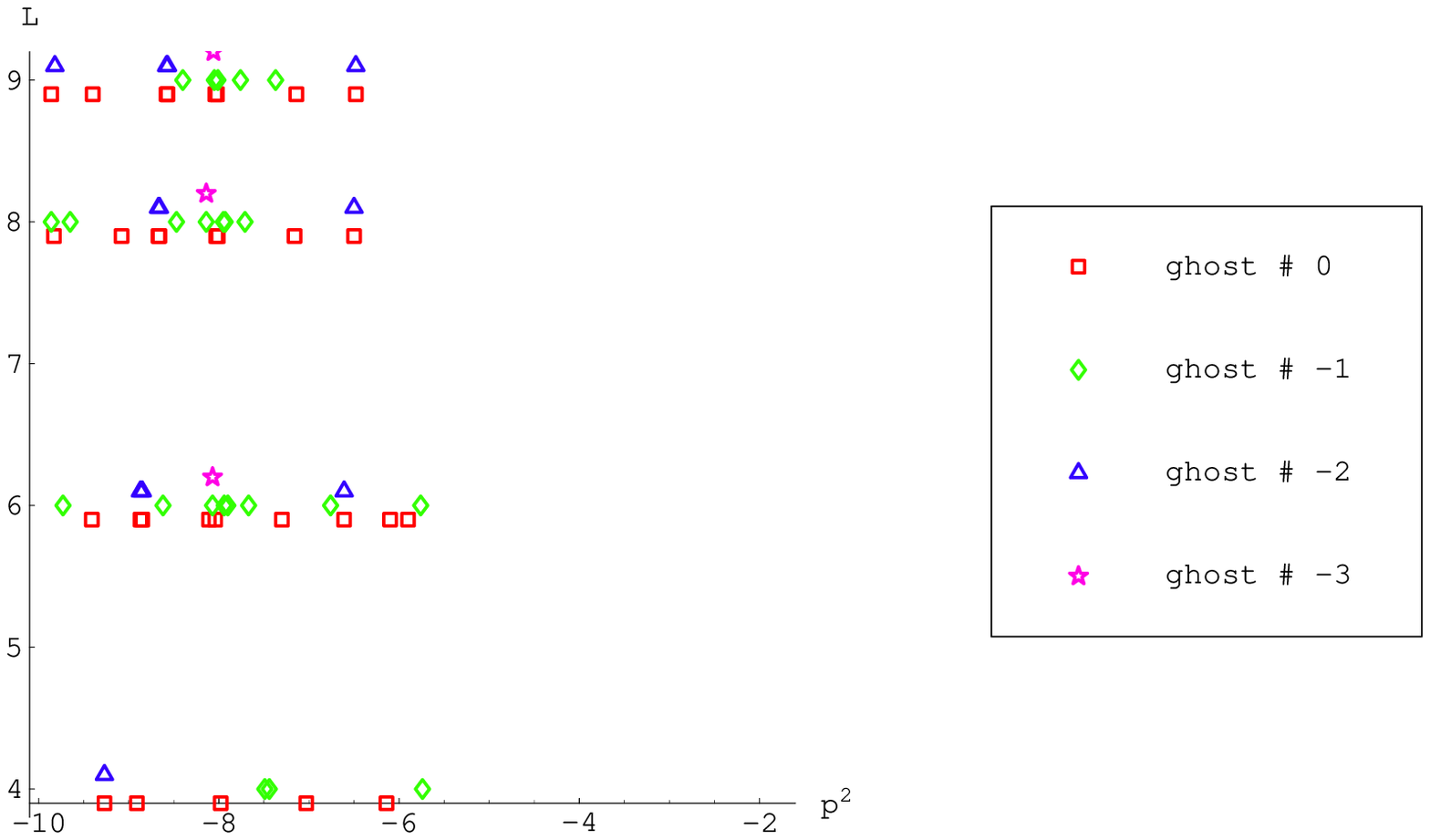}
\end{center}
\caption[x] {\footnotesize Zeros of $\Delta^\ss{n}_{-}(p^2)$ (a) and of
$\Delta^\ss{n}_{+}(p^2)$ (b) for $n=0,-1,-2$ 
at levels $L= 4,\ldots,9$ up to $p^2 =-10$. }
\label{f:nextzeroes_gaugefixed.eps}
\end{figure}

\noindent All the determinants 
\be
\Delta^\ss{n}_\pm(p^2) \equiv \det \Ltilde^\ss{n,\pm}(p) 
\ee 
evaluated at levels $L=4,\ldots,9$ do not vanish for $p^2\ge 0$. On general
grounds we expect that the LT approximation is meaningful
for $p^2 \gg -2L$. We observe indeed that only the zeros of the
determinants that are sufficiently close to $p^2=0$ are stable when
the level increases: for example, in the {\it odd} twist sector the first
group of zeros, shown in Figure~\ref{f:zeroes_gaugefixed.eps},
is centered around $p^2 \approx -2$ and it is quite stable as $L$ goes
from 4 to 9.  The next group of zeros in the odd twist sector
appears around $p^2\approx -6$
and in this region not all the zeros are yet stabilized for $L=9$ 
(Figure~\ref{f:nextzeroes_gaugefixed.eps} (a)): there is multiplet
of zeros that has vanishing index for levels 5 and 6 but when going to
levels 7, 8 and 9 a {\it pair} of zeros of $\Delta^\ss{-1}_{-}$ disappears
making the index jump to 4. This pair of zeros corresponds
to a single eigenvalue of $\Ltilde^\ss{-1}$ that has two almost 
coincident zeros at $p^2\approx -6$ for
levels 5, 6 which become a pair of complex conjugate zeros with
a small imaginary part at higher levels. It is thus not unlikely
that this pair would come back on the real $p^2$ axis if the computation were
pushed to levels higher than 9.
 
In the {\it even} sector the first zeros of $\Delta^\ss{n}_{+}$  
show up for $p^2 \approx -6$: in this region the zeros of the determinants 
are definitely not stable in the range of levels that we were able to
probe, there is no clear multiplet structure that we can detect 
and the approximation does not appear to be reliable   
(Figure~\ref{f:nextzeroes_gaugefixed.eps} (b)).
In conclusion in the region where it appears that the LT approximation
is relatively accurate ($p^2\gtrsim -5$) zeros of the determinants only 
occur in the odd sector: 
all these zeros are located around $p^2 \approx -2$ and  well separated
from any other zeros. Their center of mass is at 
\be
{\bar m}^2 =-2.32, -2.22, -2.19, -2.16, -2.14, -2.13 \quad {\rm for}\quad  L=4,\ldots,9
\ee 
and their spreads 
\be
\Delta m^2= 0.45, 0.27, 0.29, 0.19, 0.22, 0.16 \qquad {\rm for}\quad L=4,\ldots,9
\ee
tend to decrease as the level goes up\footnote{Actually $\Delta m^2$ increases slightly
when going from $L=7$ to $L=8$ and then decreases again for $L=9$. This can be
explained by the fact that the approximation is of type $(L,3L)$ for $L=7$ 
and $(L,2L)$ for $L=8, 9$.}.

Therefore it is reasonable to conclude that 
they correspond to a single determinant
multiplet. For this group of zeros the Fadeev-Popov index (\ref{fpindex}) does
vanish: 
\be
I_{FP}({\bar m}) = d_0 - 2\, d_1 + 2\, d_2 = 2- 2\cdot 2+ 2\cdot 1 =0
\label{fpindexmbar}
\ee
in agreement with Sen's conjecture. Note also that the zero of
$\Delta^{(-2)}_{-}$ is exactly degenerate with one of the two zeros of
$\Delta^{(0)}_{-}$. This is due to the $SU(1,1)$ symmetry we mentioned
above. Therefore it is not possible to split the multiplet in
Figure~\ref{f:zeroes_gaugefixed.eps} into two or
more smaller multiplets with non-negative indices: the only way to get
a non-negative index is to consider all the zeros as corresponding to
degenerate fields carrying no total number of physical degrees of freedom.

\section{Relative and Absolute Cohomologies}
We already recalled that the number of physical states of open string
theory is given by the dimension of the cohomology $\H0(\Qtilde)$ 
on $\F0$, the space of states of CFT ghost number 0.  
Let us also introduce the spaces $\H{n}(\Qtilde)$, the
$\Qtilde$-cohomologies on $\F{n}$, the states of CFT ghost number 
$n\not=0$: in the following we will refer to
the $\H{n}(\Qtilde)$'s as the {\it absolute} $\Qtilde$-cohomologies.
Since $\Qtilde$ is symmetric with respect to the
non-degenerate bilinear form $(\,\cdot\,,\,\cdot)$, 
the following duality between absolute cohomologies holds
\be
\H{n}(\Qtilde) = \H{1-n}(\Qtilde)
\label{absolutedual}
\ee
In the perturbative case, the absolute cohomologies which are 
non-vanishing (for non-exceptional momenta) are $\H0(Q)$
 --- the  physical state cohomology --- and
its dual $\H1(Q)$. In this section
we will probe the non-perturbative cohomologies 
$\H{n}(\Qtilde)$ for $n = -1,-2$ (and their duals)
and we will exhibit evidence for their non-emptiness.

One way to compute $\H{n}(\Qtilde)$ is based on the preliminary
computation of a different kind of $\Qtilde$-cohomologies --- the   
{\it relative} cohomologies. The  $\Qtilde$-cohomology of ghost number $n$
{\it relative to $b_0$} 
is  defined on the space $\Wtilde{n}$ of states $\phi_n$ of ghost number $n$  
which are $b_0$ and $\Ltilde$ invariant:
\be
\phi_n\in \Wtilde{n}\,  
\mathrel{\mathop{\iff}^{\rm def}}\, b_0\,\phi_n =\Ltilde\,\phi_n =0
\ee
The  relative $\Qtilde$-cohomology of ghost number $n$ is given by the 
$\Qtilde$-closed  states $\phi_n\in \Wtilde{n}$
\be
\Qtilde\, \phi_n =0
\ee
modulo the states which are in the $\Qtilde$ image of $\Wtilde{n-1}$
\be
\phi_{\sss n} \sim \phi^\prime_{\sss n} =
\phi_{\sss n} + \Qtilde\, \phi_{\sss n-1}
\ee
where $\phi_{\sss n-1}\in \Wtilde{n-1}$. Such a definition is
consistent since 
\be
\{\Qtilde, b_0\} = \Ltilde
\label{dueltilde}
\ee
The relative cohomologies of
$\Qtilde$ will be denoted by $\htilde{n}$.

To unravel the relation between absolute and relative $\Qtilde$ cohomologies
it is useful to review the same relation for the perturbative $Q$. 
$Q$ can be decomposed as follows
\be
Q = c_0\, L_0 + b_0\, D + M 
\label{pertdecomposition}
\ee
where $L_0$, $D$ and $M$ are independent of $c_0$ and $b_0$. $Q^2=0$
implies the relations
\be
M^2 + D\,L_0 =0 \qquad [D, L_0] =[L_0, M] = [M, D] = 0 
\label{nilpert}
\ee
Note that the first of the equations above implies that $M^2=0$ on
$\W{n}$, the space of $b_0$ and $L_0$ invariant states: 
thus the cohomology of $Q$ relative to $b_0$, $\h{n}$, coincides with the
cohomology of $M$ on $\W{n}$.
Since $\{ Q, b_0\}=L_0$, $Q$-closed states which are not $L_0$-invariant 
are necessarily $Q$-trivial. Therefore the computation of the 
absolute cohomology
can be restricted, with no loss of generality, 
to the subspace $\V{n}$ of the open string state space with
$L_0=0$. Define the following maps
\bea
&\!\!\!\!\!\!\!\!\!\imath\, : \W{n} \rightarrow \V{n}\qquad &\imath(\phi_n) =\phi_n\nonumber\\
&\!\!\!\!\!\!\pi\, : \V{n} \rightarrow \W{n-1}\qquad &\pi(\phi_n+c_0\phi_{n-1}) 
=\phi_{n-1}\nonumber\\
&{\cal D}\, : \W{n-1} \rightarrow \W{n+1}\qquad &{\cal D}(\phi_{n-1}) = D\,\phi_{n-1}
\label{pertbott}
\eea
It is simple to check that $Q\, \imath = \imath\, M$,
$M\, \pi = -\pi\, Q$ and $M\, D = D\,M$. Therefore
$\imath$, $\pi$ and ${\cal D}$ descend to cohomology maps:
\be
\cdots\; {\mathop{\longrightarrow}^{\cal D}}\; \h{n}\;
{\mathop{\longrightarrow}^{\imath }}\;\H{n}(Q)\; {\mathop{\longrightarrow}^{\pi}}
\;\h{n-1}\;{\mathop{\longrightarrow}^{\cal D}}\;\h{n+1}\;
{\mathop{\longrightarrow}^{\imath}}\;\cdots
\label{pertbottsequence}
\ee
It is straightforward to verify  
that the sequence of maps above defines a cohomology complex,
\be
\imath\,{\cal D} = \pi\,\imath ={\cal D}\, \pi =0
\ee
and that moreover the cohomology of this complex is trivial:
\be
{\rm img}\, {\cal D} =\ker\, \imath \qquad {\rm img}\, 
\imath = \ker\, \pi\qquad {\rm img}\,\pi= \ker\, {\cal D}
\ee
The exact long sequence (\ref{pertbottsequence}) 
describes the perturbative absolute cohomologies in terms of the 
relative ones. This
is useful since one can establish by other methods that 
\be
\h{n}= \delta_{n,0}\, \h0
\ee
at non-exceptional momenta. Then the exact long sequence 
(\ref{pertbottsequence}) breaks 
into short ones:
\bea
&&0 = \h{-2}\;{\mathop{\longrightarrow}^{\cal D}}\; \h{0}\;
{\mathop{\longrightarrow}^{\imath }}\;\H{0}(Q)\; {\mathop{\longrightarrow}^{\pi}}\;\h{-1} = 0\nonumber\\
&&0 = \h{n}\;{\mathop{\longrightarrow}^{\imath }}\;\H{n}(Q)\; 
{\mathop{\longrightarrow}^{\pi}}\;\h{n-1}=0\qquad {\rm if}\quad n\not=0,1
\eea
One proves in this way that 
\be
\H{0}(Q) \simeq \H{1}(Q)\simeq \h{0} \qquad{\rm and}\qquad  \H{n}(Q)=0
\quad {\rm if}\quad n\not=0,1
\ee

Our goal in the rest of this subsection will be  
to investigate the relation between the non-perturbative $\htilde{n}$
and $\H{n}(\Qtilde)$ along similar lines and to write down
the generalization of the long exact sequence (\ref{pertbottsequence}). 
We begin by decomposing $\Qtilde$ in terms of $b_0$ and $c_0$:
\be
\Qtilde = c_0\, \Lhat + b_0\,\Dhat + \Mhat  + c_0\, b_0\, \Zhat
\label{nonpertdecomposition}
\ee
where $\Lhat$, $\Dhat$, $\Mhat$ and $\Zhat$ are independent of $c_0$
and $b_0$.  The crucial difference between the decomposition 
(\ref{nonpertdecomposition}) of the non-perturbative $\Qtilde$ and 
its perturbative analogue (\ref{pertdecomposition}) is the term proportional
to $c_0\, b_0$, which is absent in the perturbative case. Note that
\be
\Ltilde \equiv \{\Qtilde, b_0\} = \Lhat + b_0\, \Zhat 
\qquad \Dtilde \equiv \{\Qtilde, c_0\} = \Dhat - c_0\,\Zhat
\label{zed}
\ee
and therefore $[\Ltilde, c_0] = [b_0,\Dtilde]= -\Zhat$, in agreement with the
Jacobi identity (\ref{iacobi}). 

It is worth pausing here to remark that the $(b_0,c_0)$ expansion of the 
first quantized BRS operator $\Qtilde_{bcft}$ associated with a generic
boundary matter conformal field theory coupled to 2d gravity has
$\Zhat=0$. However even if the non-perturbative tachyonic vacuum of OSFT
were described by such a boundary conformal field theory this would
not mean, necessarily, that $\Zhat =0$ in the expansion
(\ref{nonpertdecomposition}) for $\Qtilde$; it would only imply that
$\Qtilde$ is {\it conjugate}, by means of a linear field redefinition $U$, to
an operator $\Qtilde_{bcft}$ whose $(b_0, c_0)$ expansion has $\Zhat
=0$. {\it If} $U$ commuted with $b_0$, the relative complex $(\Qtilde,
b_0)$ would be equivalent to the complex $(\Qtilde_{bcft},
b_0)$ (and in this case $\Zhat = [\Lhat, X]$): 
however, in general, we do not know if the field
redefinition $U$ that eliminates $\Zhat$ also commutes with $b_0$.
Since we are looking for properties of the relative complex $(\Qtilde,
b_0)$ we must consider the general case in which $\Zhat$ is
non-trivial. In the following we will elucidate the complications
that a non-vanishing $\Zhat$ entails for the relationship between 
absolute and relative  cohomologies of $\Qtilde$.

The nilpotency of $\Qtilde$ leads
to equations that replace the perturbative ones 
(\ref{nilpert}):
\bea
&&\Mhat^2 + \Dhat\,\Lhat =0 \qquad \{\Mhat, \Zhat\} + \Zhat^2 = [\Dhat, \,\Lhat] 
\nonumber\\
&&\Lhat\, \Mhat - (\Mhat +\Zhat)\, \Lhat =0\qquad
\Mhat\,\Dhat -\Dhat\, (\Mhat +\Zhat) =0
\label{nilnonpert}
\eea
These equations show that, like in the perturbative case, the
$b_0$-relative cohomology $\htilde{n}$ is the cohomology of the
operator $\Mhat$ on $\Wtilde{n}$, the space of states which are $b_0$
and $\Ltilde$ invariant: indeed, the first of the equations
(\ref{nilnonpert}) says that $\Mhat^2=0$ on $\Wtilde{n}$, since
Eq. (\ref{zed}) ensures that $\phi_n\in\Wtilde{n}\Rightarrow
\Lhat\,\phi_n=0$.  Moreover the third of the equations (\ref{nilnonpert}) 
guarantee that $\Mhat: \Wtilde{n}\to\Wtilde{n+1}$. Let us denote by 
$\Vtilde{n}$ the space of the
states which are $\Ltilde$ invariant: 
\be
(\phi_n+ c_0\,\phi_{n-1})\in \Vtilde{n}\,  
\mathrel{\mathop{\iff}^{\rm def}}\,
\Ltilde\,(\phi_n+c_0\,\phi_{n-1}) =0
\ee
In what follows we will assume that, analogously to the perturbative case, 
the open string state space of ghost number $n$ decomposes as follows
\be
\F{n} = \Vtilde{n}\oplus {\rm img}(\F{n}; \Ltilde)
\label{Fdecomp}
\ee
where ${\rm img}(\F{n}; \Ltilde)$ is the image of $\F{n}$ under the map 
$\Ltilde$. The decomposition above 
would follow from the symmetry of $\Ltilde$ with respect to a positive
definite bilinear form. However $(\cdot\,,\,\cdot)$ is not positive
definite and thus (\ref{Fdecomp}) appears to be an independent
hypothesis. (\ref{Fdecomp}) ensures that, as in the perturbative
case, the absolute $\Qtilde$-cohomology is contained in the
kernel of $\Ltilde$, $\Vtilde{n}$.

Let us introduce the immersion and projection maps $\imath$ and $\pi$:
\bea
&\!\!\!\!\!\!\!\!\!\imath\, : \Wtilde{n} \rightarrow \Vtilde{n}\qquad &\imath(\phi_n) =\phi_n\nonumber\\
&\!\!\!\!\!\!\pi\, : \Vtilde{n} \rightarrow \Wcheck{n-1}\qquad &
\pi(\phi_n+c_0\phi_{n-1}) 
=\phi_{n-1}
\label{nonpertbott}
\eea
where $\Wcheck{n}$ is the subspace of $\Wtilde{n}$ defined as follows
\be
\phi_n\in \Wcheck{n}\,  
\mathrel{\mathop{\iff}^{\rm def}}\, \phi_n\in\Wtilde{n},\,\, \Zhat\, \phi_n = \Lhat\, \phi_{n+1}, \,\, \phi_{n+1}\in \O{n+1} 
\label{checkdef}
\ee
On the space $\O{n}$ of $b_0$-invariant states, the kinetic operator $\Ltilde$
of the gauge-fixed open string field theory reduces to $\Lhat$. 
Since $b_0$ commutes with $\Ltilde$, the decomposition 
(\ref{Fdecomp}) of the total open string state space $\F{n}$
induces the following decomposition of $\O{n}$ 
as the sum of a vector in $\Wtilde{n}$, the kernel 
of $\Lhat$, and a vector in the  image of $\Lhat$ in $\O{n}$:
\be
\O{n} = \Wtilde{n} \oplus {\rm img}(\O{n}; \Lhat)
\label{omegadecomp}
\ee
Correspondingly one can write the following decomposition for $\Zhat$:
\be
\Zhat = \Zcheck + \Lhat\, X
\label{zeddecomp}
\ee
where $\Lhat\, \Zcheck =0$, and the operator $X : \O{n}\to \O{n+1}$ 
is defined up to
an operator whose image is in the kernel of $\Lhat$. Therefore the space
$\Wcheck{n}$ defined in Eq. (\ref{checkdef}) coincides with the
kernel of $\Zcheck$ in $\Wtilde{n}$:
\be
\Wcheck{n} = \ker \Zcheck \cap \Wtilde{n}
\ee 
$\imath$ and $\pi$ define the following exact short sequence:
\be
0\; \longrightarrow \Wtilde{n}\; {\mathop{\longrightarrow}^{\imath }}\;\Vtilde{n}\; 
{\mathop{\longrightarrow}^{\pi}}\;\Wcheck{n-1}\; \longrightarrow\; 0
\label{bottsequence1}
\ee
Note that $\Qtilde\, \imath = \imath\, \Mhat$ and
$\Mhat\, \pi = -\pi\, \Qtilde$. Moreover 
\be
\Mhat :\Wcheck{n} \to \Wcheck{n+1}
\label{wcheckinvariance}
\ee
by virtue of the nilpotency relations (\ref{nilnonpert}).
To see this, insert the
decomposition (\ref{zeddecomp}) of $\Zhat$ into the second equation in
(\ref{nilnonpert}):
\bea
&&\{\Mhat, \Zcheck\} + \Zcheck^2 +\{\Mhat+\Zcheck, \Lhat\, X\}
+ (\Lhat\, X)^2 =\nonumber\\
&&= \{\Mhat, \Zcheck\} + \Zcheck^2 + (\Mhat+ \Zhat)\, \Lhat\, X + 
\Lhat\, X\,(\Mhat+\Zcheck) =\nonumber\\
&&= \{\Mhat, \Zcheck\} + \Zcheck^2  + \Lhat\, (\{\Mhat, X\} +X\,\Zcheck) =[\Dhat, \,\Lhat] 
\eea
where we used the the third equation in  (\ref{nilnonpert}). 
Applying both sides of this  equation to $\Wtilde{n-2}$ and
decomposing their image in $\O{n}$ according to (\ref{omegadecomp})
one obtains 
\be
\{\Mhat, \Zcheck\} + \Zcheck^2 =0\qquad \Lhat\, (\{\Mhat, X\} +X\,\Zcheck +\Dhat) = 0
\qquad {\rm on}\quad\Wtilde{n}
\label{curvaturedecomp}
\ee
The first of these relations implies that $\Mhat$ and $\Zcheck$ anti-commutes
on $\Wcheck{n}$, thus Eq. (\ref{wcheckinvariance}) holds.

In conclusion, the following diagram is (anti)-commutative

\be
\def\normalbaselines{\baselineskip20pt \lineskip3pt\lineskiplimit3pt}
\def\mapright#1{\smash{\mathop{\longrightarrow}\limits^{#1}}}
\def\mapdown#1{\Big\downarrow\rlap{$\vcenter{\hbox{$\scriptstyle#1$}}$}}
\matrix{
0&\mapright{}&\Wtilde{n}&\mapright\imath&\Vtilde{n}
&\mapright\pi&\Wcheck{n-1}&\mapright{}&0\cr
&&\mapdown\Mhat&&\mapdown\Qtilde&&\mapdown\Mhat\cr
0&\mapright{}&\Wtilde{n+1}&\mapright\imath&\Vtilde{n+1}
&\mapright\pi&\Wcheck{n}&\mapright{}&0\cr
}
\label{commutativediagram}
\ee

\noindent We are thus in condition to apply the general theorem
of \cite{bott}: the short sequence (\ref{bottsequence1})
gives rise to the following {\it exact} long sequence of 
$\Qtilde$-cohomologies
\be
\cdots\; {\mathop{\longrightarrow}^{\Dcaltilde}}\; \htilde{n}\;
{\mathop{\longrightarrow}^{\imath }}\;\H{n}(\Qtilde)\; {\mathop{\longrightarrow}^{\pi}}
\;\hcheck{n-1}\;{\mathop{\longrightarrow}^{\Dcaltilde}}\;\htilde{n+1}\;
{\mathop{\longrightarrow}^{\imath}}\;\cdots
\label{nonpertbottsequence}
\ee
This is the non-perturbative generalization of the sequence 
(\ref{pertbottsequence}) that we were seeking for. $\Dcaltilde$ is the operator
\be
\Dcaltilde \equiv \Dhat + \Mhat\, X
\ee
and $\hcheck{n}$ is a new kind of cohomology, 
the cohomology of $\Mhat$ on $\Wcheck{n}$, whose representatives
$\phi_n$ satisfy the following conditions
\be 
\Mhat\, \phi_n =0, \; \phi_n \in \Wcheck{n},\quad \phi_n \sim \phi_n +
\Mhat\, \phi_{n-1} ,\; \phi_{n-1} \in \Wcheck{n-1}
\ee
$\hcheck{n}$ is the same as the cohomology of $\Mhat$ relative
to $\Zcheck$.  It is well-defined
since $\Mhat$ sends the kernel of $\Zcheck$ into itself (See Eq. 
(\ref{wcheckinvariance})). When $\Zcheck\not= 0$, $\hcheck{n}$
differs, in general, from the relative cohomology $\htilde{n}$. Therefore,
in presence of a non-vanishing $\Zcheck$, one cannot
express the absolute $\Qtilde$-cohomology only in terms of the relative
cohomologies $\htilde{n}$ --- one needs the knowledge of the cohomologies
$\hcheck{n}$ as well. 
We remarked earlier that when $\Qtilde$ is conjugate to a BRS operator 
with $\Zhat=0$ by means of a field redefinition which preserves $b_0$, $\Zhat$
must be a $\Lhat$-commutator. In this case then $\Zcheck =0$ on $\Wtilde{n}$,
and $\hcheck{n}=\htilde{n}$.
 
\subsection{Relations between relative cohomologies}

In this subsection we want to investigate the relation between 
the cohomologies $\hcheck{n}$ and $\htilde{n}$ that appear in the 
non-perturbative long exact sequence (\ref{nonpertbottsequence}). This 
relation will be expressed by the two long exact sequences 
that are written in Eq. (\ref{oursequences}) below.

There is an obvious immersion $\iota_{\sss 1}: \hcheck{n}\rightarrow
\htilde{n}$ of $\hcheck{n}$ into $\htilde{n}$, given by the identity
map. In general this immersion is neither injective nor
surjective. The kernel of the immersion is represented by vectors
$\phi_n$ which are trivial in $\htilde{n}$ but not in $\hcheck{n}$:
\be
\phi_n = \Mhat\,\phi_{n-1},\; {\rm with}\; \Zcheck\phi_{n-1}\not=0\quad
{\rm and}\;\Zcheck\phi_n=0
\ee
The cokernel is given by the $\Mhat$-closed $\phi_n$ which are not
$\Zcheck$-invariant. 

On $\Wtilde{n}$ there exists another nilpotent operator beyond
$\Mhat$: indeed, the first of the relations (\ref{curvaturedecomp}) implies 
that the operator $\Mhat +\Zcheck$ is nilpotent on $\Wtilde{n}$. 
Let us denote the cohomology of $\Mhat +\Zcheck$ on
$\Wtilde{n}$ by $\H{n}(\Mhat+\Zcheck)$. 

The existence of the non-degenerate bilinear form
$\langle\cdot\,,\,\cdot\rangle$ on $\O{-n}\times\O{n}$ ensures that
the kernels $\Wtilde{n}$ of the operators
$\Lhat$ satisfies the following duality relation 
\be 
\Wtilde{n} = \Wtilde{-n}
\label{nduality}
\ee 
The decomposition (\ref{omegadecomp}) 
guarantees that $\langle\cdot\,,\,\cdot\rangle$
is non-degenerate on $\Wtilde{-n}\times\Wtilde{n}$.
The symmetry of $\Qtilde$ with respect to the
bilinear form $(\,\cdot\,,\,\cdot\,)$ is equivalent to the relations 
\be
\Lhat^\dagger =\Lhat \qquad \Dhat^\dagger =\Dhat\qquad \Mhat^\dagger =
\Mhat +\Zhat\qquad \Zhat^\dagger =-\Zhat
\ee
where the dagger denotes the adjoint conjugation with respect
to the bilinear form $\langle\cdot\,,\,\cdot\rangle$.
Therefore if $\Zhat\not=0$, $\Mhat$ fails to be symmetric: 
the adjoint of $\Mhat$ with respect to the
bilinear form $\langle\cdot\,,\,\cdot\rangle$ on $\O{-n}
\times\O{n}$ is $\Mhat +\Zhat$ and
the adjoint of $\Mhat$ with respect to the
bilinear form $\langle\cdot\,,\,\cdot\rangle$ on $\Wtilde{-n}
\times\Wtilde{n}$ is $\Mhat +\Zcheck$.
Therefore, thanks to Hodge decomposition,
the cohomology of $\Mhat$ at ghost number $n$ is isomorphic 
to the cohomology of $\Mhat +\Zcheck$ at ghost number $-n$: 
\be 
\H{n}(\Mhat+\Zcheck) =\htilde{-n} 
\ee
In the perturbative case, when $\Zcheck =0$, the relation above
reduces to the duality between relative cohomology $\h{n}=\h{-n}$.

The identity map provides an immersion $\iota_{\sss 2}$ of the 
cohomology $\hcheck{n}$ into $\H{n}(\Mhat+\Zcheck)$. Again,
in general, $\iota_{\sss 2}$ is neither injective nor surjective. The
kernel of $\iota_{\sss 2}$ is represented by vectors $\phi_n =(\Mhat+\Zcheck)\,\phi_{n-1}$ with $\Mhat\, \Zcheck\,\phi_{n-1}=0$ and $\Zcheck\,\phi_{n-1}\not=0
$. The cokernel is represented by vectors $\phi_n$ with $(\Mhat+\Zcheck)\,\phi_{n}=0$ and $\Mhat\,\phi_{n}\not=0$.

In the following we will derive two exact long sequences
of cohomologies
which captures the lack of injectivity and surjectivity of the
immersions $\iota_{\sss 1}$ and $\iota_{\sss 2}$.

Consider the short exact sequence between vector spaces
\be
0\; \longrightarrow \Wcheck{n}\;{\mathop{\longrightarrow}^{\rm Id}}
\;\Wtilde{n}\;{\mathop{\longrightarrow}^{\Zcheck}}\; {\rm img}
(\Wtilde{n};\Zcheck)\;\longrightarrow\;0
\label{exactshort2}
\ee
where  ${\rm img}(\Wtilde{n};\Zcheck)$ is the image of 
$\Wtilde{n}$ under the map $\Zcheck$.  Both $\Mhat$ and $\Mhat+\Zcheck$
map  ${\rm img}(\Wtilde{n};\Zcheck)$ into  ${\rm img}(\Wtilde{n+1};\Zcheck)$:
\bea
&&\Mhat\, \Zcheck \phi_n = -\Zcheck\, (\Mhat +\Zcheck)\phi_n \in 
{\rm img}(\Wtilde{n+1};\Zcheck)\nonumber\\
&&(\Mhat+\Zcheck)\, \Zcheck \phi_n = -\Zcheck\, \Mhat \phi_n \in 
{\rm img}(\Wtilde{n+1};\Zcheck)
\eea
Therefore the cohomologies of both $\Mhat$ and $\Mhat+\Zcheck$ are
well-defined on the spaces 
${\rm img}(\Wtilde{n-1};\Zcheck)$: these cohomologies
will be denoted with $\hhat{n}_{\sss 1}$ and $\hhat{n}_{\sss 2}$,
respectively.

Thus, on the vector spaces
$(\Wcheck{n},\Wtilde{n}, {\rm img}(\Wtilde{n};\Zcheck))$ appearing in the 
short sequence (\ref{exactshort2})
we can consider either the coboundary operators 
$(\Mhat, \Mhat, \Mhat+\Zcheck)$ or the operators $(\Mhat, \Mhat+\Zcheck, 
\Mhat)$. Both triples, together with the short sequence (\ref{exactshort2}), 
give rise to (anti)-commutative diagrams like that in 
(\ref{commutativediagram}). One concludes that the following two long
sequences of cohomologies are exact:
\bea
&&\cdots\;{\mathop{\longrightarrow}^{\sss \Mhat\Zcheck^{-1}}}\; 
\hcheck{n}\;
{\mathop{\longrightarrow}^{\iota_{\sss 1}}}
\;\htilde{n}\; {\mathop{\longrightarrow}^{\sss\Zcheck}}
\;\hhat{n+1}_{\sss 1}\;{\mathop{\longrightarrow}^{\sss\Mhat\Zcheck^{-1}}}\;
\hcheck{n+1}\;{\mathop{\longrightarrow}^{\iota_{\sss 1}}}\;\cdots\nonumber\\
&&\cdots\; {\mathop{\longrightarrow}^{\sss 1+\Mhat\Zcheck^{-1}}}\; \hcheck{n}\;
{\mathop{\longrightarrow}^{\iota_{\sss 2}}}
\;\htilde{-n}\; {\mathop{\longrightarrow}^{\sss \Zcheck}}
\;\hhat{n+1}_{\sss 2}\;{\mathop{\longrightarrow}^{\sss 1+\Mhat\Zcheck^{-1}}}\;
\hcheck{n+1}\;{\mathop{\longrightarrow}^{\iota_{\sss 2}}}\;\cdots
\label{oursequences}
\eea

Thus we see that $\htilde{n} =\hcheck{n}$ only if $\hhat{n}_{\sss 1}=
\hhat{n+1}_{\sss 1}=0$ and $\htilde{n} =\hcheck{-n}$ only if
$\hhat{n}_{\sss 2}=\hhat{n+1}_{\sss 2}=0$. 

Let us remark that  
$\hhat{n}_{\sss 1}$ and $\hhat{n}_{\sss 2}$ satisfy simple duality relations:
\be
\hhat{n}_{\sss 1} =\hhat{1-n}_{\sss 1}\qquad 
\hhat{n}_{\sss 2} =\hhat{1-n}_{\sss 2}
\label{poincare}
\ee
Indeed, the non-degenerate bilinear form $\langle\cdot\,,\,\cdot\rangle$ on
$\Wtilde{-n}\times\Wtilde{n}$ projects to a non-degenerate bilinear
form $\langle\!\langle\cdot\,,\,\cdot\rangle\!\rangle$ on
${\rm img}\, (\Wtilde{n-1};\Zcheck)\times{\rm img}\, (\Wtilde{-n};\Zcheck)$
defined by
\be
\langle\!\langle\phi_{n} ,\, \phi_{-n+1}\rangle\!\rangle\equiv
\langle\phi_{n},\,\Zcheck^{-1}\phi_{-n+1}\rangle 
\ee
Such a non-degenerate form provides the identification 
\be
{\rm img}\, (\Wtilde{n-1};\Zcheck) =  {\rm img}\, (\Wtilde{-n};\Zcheck)
\ee
With respect to $\langle\!\langle\cdot\,,\,\cdot\rangle\!\rangle$, both the
operators $\Mhat$ and $\Mhat+\Zcheck$ are antisymmetric:
\bea
&&\langle\!\langle\Mhat\phi_{n},\,\phi_{-n}\rangle\!\rangle
= \langle\Mhat\phi_{n},\,\Zcheck^{-1}\phi_{-n}\rangle = \langle \phi_{n},\,(\Mhat+\Zcheck)\,\Zcheck^{-1} \phi_{-n}\rangle=\nonumber\\
&& = -\langle \phi_{n},\,\Zcheck^{-1}\,\Mhat \phi_{-n}\rangle=
- \langle\!\langle \phi_{n},\,\Mhat\,\phi_{-n}\rangle\!\rangle
\eea
and similarly for $\Mhat+\Zcheck$. Thus the duality relations (\ref{poincare})
follow by virtue of Hodge decomposition. 
\subsection{The numerical data}

In this  subsection we will start from the numerical computation
of the dimensions of $\Wtilde{n}$ in the region of negative $p^2$ where the LT
approximation appears to be reliable. We will explain that our numerical 
data imply that
for $p^2 =-{\bar m}^2$ a certain linear combination of dimensions of
relative cohomologies $\htilde{n}$ (in the odd twist parity sector) 
is non-vanishing. From $\H0(\Qtilde)=0$ at $p^2=-{\bar m}^2$
and from the long exact sequence (\ref{nonpertbottsequence}), we will derive 
some linear
relations for the absolute cohomologies at ghost numbers $n=-1,-2$
which involve also the relative cohomologies $\htilde{n}$ and $\hcheck{n}$.
These relations cannot be satisfied by $\H{-2}(\Qtilde)=\H{-1}(\Qtilde)=0$, 
since not all the $\htilde{n}$ can vanish.
By imposing the constraints among
$\htilde{n}$, $\hcheck{n}$ and $\hhat{n}_{\sss 1,2}$ which follow from
the sequences (\ref{oursequences}) and the duality relations (\ref{poincare})
derived in the previous subsection,
we will be able to determine all possible values for
the dimensions of the various cohomologies that entered our
discussion. It will turn out that only 3 different solutions
for the dimensions of $\htilde{n}$, $\hcheck{n}$ and $\hhat{n}_{\sss 1,2}$
are compatible both with the ``experimental'' fact that a 
certain linear combination  of dimensions of $\htilde{n}$ must not vanish and
with the constraints that descend from the sequences 
(\ref{nonpertbottsequence}), (\ref{oursequences}), (\ref{poincare}) --- 
and for all three of them $\H{-2}(\Qtilde)=\H{-1}(\Qtilde)=1$.

We saw in the previous section that the numerical computation is consistent
with Sen's conjecture about the vanishing of  $\H0(\Qtilde)$ for all values 
of $p^2$. Assuming then  $\H0(\Qtilde)= \H1(\Qtilde) = 0$, the long sequence
(\ref{nonpertbottsequence}) breaks into the short exact sequence
\be
0\; {\mathop{\longrightarrow}^{\pi}}
\;\hcheck{-1}\;{\mathop{\longrightarrow}^{\Dcaltilde}}\;\htilde{1}\;
{\mathop{\longrightarrow}^{\imath}}\; 0
\label{bottshort}
\ee
and into the  two semi-infinite exact sequences
\bea
&&\!\!\!\!\!\!\!\!\cdots\;{\mathop{\longrightarrow}^{\pi}}\;\hcheck{-3}\; 
{\mathop{\longrightarrow}^{\Dcaltilde}}\; \htilde{-1}\;
{\mathop{\longrightarrow}^{\imath }}\;\H{-1}(\Qtilde)\; {\mathop{\longrightarrow}^{\pi}}
\;\hcheck{-2}\;{\mathop{\longrightarrow}^{\Dcaltilde}}\;\htilde{0}\;
{\mathop{\longrightarrow}^{\imath}}\;0\nonumber\\
&&\!\!\!0\;{\mathop{\longrightarrow}^{\pi}}\;\hcheck{0}\; 
{\mathop{\longrightarrow}^{\Dcaltilde}}\; \htilde{2}\;
{\mathop{\longrightarrow}^{\imath }}\;\H{-1}(\Qtilde)\; {\mathop{\longrightarrow}^{\pi}}
\;\hcheck{1}\;{\mathop{\longrightarrow}^{\Dcaltilde}}\;\htilde{3}\;
{\mathop{\longrightarrow}^{\imath}}\;\cdots\nonumber\\
\label{semibottsequence1}
\eea
From (\ref{bottshort}) one obtains
\be
\htilde{1}=\hcheck{-1}
\label{relation1}
\ee
and this should hold for {\it any} $p^2$ --- if Sen's conjecture
is true.

In the region $p^2\gtrsim -5$ more detailed information about the 
cohomologies appearing in the sequences above comes from the LT numerical
computation presented in the previous Section.

Let us denote with $\Wtilde{n}^{\sss \pm}$ the kernels 
of the matrices $\Ltilde =\Lhat $ on the spaces 
$\O{n}^{\sss (\pm)}(p)$ with even and odd twist parity. 
For $p^2\gtrsim -5$, the even parity spaces 
$\Wtilde{n}^{\sss (+)} =\Wcheck{n}^{\sss (+)}$ vanish for all
$n$ and, hence, so do the even parity relative cohomologies 
\be
\htilde{n}_{\sss +} = \hcheck{n}_{\sss +}=0 \qquad \forall\; n
\ee 
From the sequences (\ref{semibottsequence1}) it follows
that the absolute cohomologies in the even twist sector 
$\H{n}_{\sss +}(\Qtilde)$ vanish for all ghost numbers:
\be
\H{n}_{\sss +}(\Qtilde)=0 \quad\forall\; n\quad {\rm and}\quad p^2 \gtrsim -5
\ee

\begin{figure}
\begin{center}
\includegraphics*[scale=.8, clip=false]{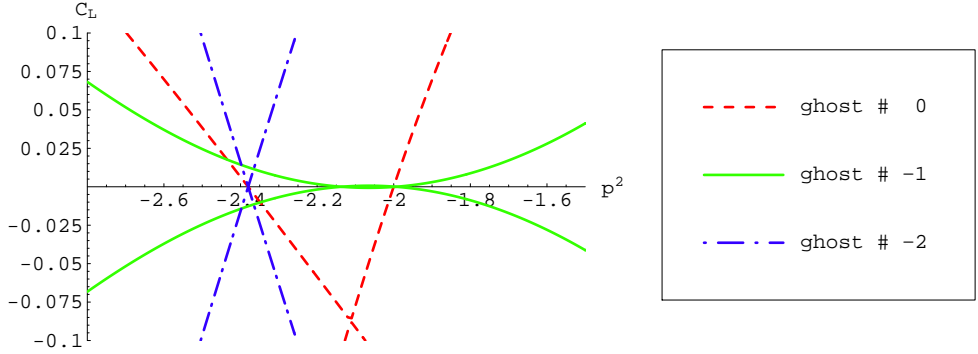}
\end{center}
\caption[x] {\footnotesize The vanishing eigenvalues of $C^{\sss (n,-)}_L(p)$
for $n=0,-1,-2$ and $p^2\approx -2$ at level L=9.}
\label{f:eigenvalues_gaugefixed.eps}
\end{figure}

In the odd twist parity sector the situation is more interesting. In the 
region $p^2\gtrsim -5$, there is one single
value of $p^2 =-{\bar m}^2 \approx -2.1$ for which the odd twist parity spaces
$\Wtilde{n}^{\sss (-)}$ for $n=0,\pm 1,\pm 2$ do not vanish. 
For $p^2= -{\bar m}^2$ we obtained for the determinant indices $d_n$
the values
\be
d_0 =2,\quad d_1 = 2, \quad d_2=1
\label{determinantindices}
\ee
From our numerical computation we can
determine not only the determinant indices (\ref{determinantindices})
but also the dimensions of the kernels of $\Lhat$ as functions
of $p^2$. Let $C^{\sss (n, -)}_L(p)$ be the 
kinetic quadratic forms 
(\ref{covariancen}-\ref{covariancezero}) in the odd twist
parity sector, at level L. In Figure~\ref{f:eigenvalues_gaugefixed.eps} we
plot the eigenvalues of $C^{\sss (n, -)}_9(p)$ (for $n=0,-1,-2$) 
that vanish for $p^2\approx -{\bar m}^2$. 
The eigenvalues of $C^{\sss (n, -)}(p)$ with $n\not=0$ come
in pairs $(\lambda_n(p), -\lambda_n(p))$: the vanishing of
a pair corresponds to a single null eigenstate of $\Ltilde^{\sss (n,-)}$.
Figure~\ref{f:eigenvalues_gaugefixed.eps} shows that for  $n=0$
there are two different eigenstates of $C^{\sss (0, -)}_9(p)$
whose eigenvalues vanish
at two approximately coincident values of $p^2$; for $n=-1$,
instead, the (approximate) double zero of $\Delta^{\sss (-1)}_-(p^2)$
corresponds to a single pair of eigenvalues that vanish (approximately)
quadratically.
In conclusion, the numerical data imply that for $p^2 = - {\bar m}^2$  
\bea
&&\dim \Wtilde{0}^{\sss (-)} = 2, \quad \dim \Wtilde{\pm 1}^{\sss (-)} = 1, 
\quad \dim \Wtilde{\pm 2}^{\sss (-)}= 
1\nonumber\\
&& \dim \Wtilde{\pm n}^{\sss (-)} = 0\quad {\rm for}\; n\geq 3
\label{experiment1}
\eea

\noindent Given the complex $\Mhat :\Wtilde{n}\rightarrow\Wtilde{n+1}$,
one has the following relation between the dimensions of $\Wtilde{n}$
and the dimensions of the cohomologies of $\Mhat$:
\be
\sum_{n=-\infty}^\infty (-1)^n\, \dim\,\Wtilde{n}=
\sum_{n=-\infty}^\infty (-1)^n\, \n{n}
\ee
where 
\be
\n{n} \equiv \dim \htilde{n}_{\sss -}
\ee  
Therefore our numerical finding (\ref{experiment1})
implies the relation 
\be
\n{-2}-  \n{-1} + \n{0} - \n{1}+ \n{2} =2
\label{experiment2}
\ee
Moreover the semi-infinite exact sequences (\ref{semibottsequence1}) 
break up at $p^2=-{\bar m}^2$ into finite sequences
\bea
&& 0\; 
{\mathop{\longrightarrow}^{\Dcaltilde}}\; \htilde{-1}_{\sss -}\;
{\mathop{\longrightarrow}^{\imath }}\;\H{-1}_{\sss -}(\Qtilde)\; {\mathop{\longrightarrow}^{\pi}}
\;\hcheck{-2}_{\sss -}\;{\mathop{\longrightarrow}^{\Dcaltilde}}\;
\htilde{0}_{\sss -}\;
{\mathop{\longrightarrow}^{\imath}}\;0\nonumber\\
&&0\;{\mathop{\longrightarrow}^{\pi}}\;\hcheck{0}_{\sss -}\; 
{\mathop{\longrightarrow}^{\Dcaltilde}}\; \htilde{2}_{\sss -}\;
{\mathop{\longrightarrow}^{\imath }}\;\H{-1}_{\sss -}(\Qtilde)\; {\mathop{\longrightarrow}^{\pi}}
\;\hcheck{1}_{\sss -}\;{\mathop{\longrightarrow}^{\Dcaltilde}}\;
0\nonumber\\
&&0\;{\mathop{\longrightarrow}^{\Dcaltilde}}\;\htilde{-2}_{\sss -}\;
{\mathop{\longrightarrow}^{\imath }}\;\H{-2}_{\sss -}(\Qtilde)\;{\mathop{\longrightarrow}^{\pi}}\; 0\nonumber\\
&&0\;{\mathop{\longrightarrow}^{\imath}}\;\H{-2}_{\sss -}(\Qtilde)\;
{\mathop{\longrightarrow}^{\pi}}\;\hcheck{2}_{\sss -}
\;{\mathop{\longrightarrow}^{\Dcaltilde}}\;0 
\label{experiment3}
\eea
Hence, we obtain 
\be
\H{-2}_{\sss -}(\Qtilde)=\htilde{-2}_{\sss -}=\hcheck{2}_{\sss -}
\label{experiment4}
\ee
from the last two sequences above, while the first two give
\bea
&&\dim \H{-1}_{\sss -}(\Qtilde) =
\n{-1}+
\ncheck{-2}- \n{0}\nonumber\\
&&\dim \H{-1}_{\sss -}(\Qtilde) = - \ncheck{0}+ \n{2}+\ncheck{1}
\label{experiment5}
\eea
where
\be
\ncheck{n} \equiv \dim \hcheck{n}_{\sss -}
\ee

We now want to look for solutions of the equations 
(\ref{experiment2}-\ref{experiment5}) with  
\be
\n0, \ncheck0 =0,1,2, \qquad \n1, \ncheck1, \n2, \ncheck2 =0,1
\label{ranges}
\ee
and
\be
\ncheck{-2} \leq \n{-2}\quad {\rm and} \quad \ncheck2 \geq \n2
\label{inequalities}
\ee
The last inequalities stem from the fact that $\Wtilde{-3} =\Wtilde{3}=0$.

It is interesting that these equations imply that
the absolute cohomologies $\H{n}(\Qtilde)$ cannot be vanishing for all
$n$.
Indeed, if $\H{-1}_{\sss -}(\Qtilde)=\H{-2}_{\sss -}(\Qtilde)=0$, 
the sequences (\ref{experiment3}) give
\be
\n{-1} = \ncheck{1} = \n{-2}=\ncheck2 =0,\qquad \n0=\ncheck{-2},
\qquad \n2=\ncheck0
\ee
Moreover since $\n0=\ncheck{-2}\leq \n{-2} =0$, it follows that $\n0 =0$. 
Eq. (\ref{experiment2}) then becomes 
\be
2 = -\n1+\n2
\ee
which does not admit any solution for $\n1$ and $\n2$ in the range 
(\ref{ranges}). 

To investigate the possible solutions of our relations
let us first consider all possible values of 
\be 
{\bf\tilde n}\equiv (\n{-2},\n{-1},\n{0},\n1,\n2)
\ee
in the range (\ref{ranges}) satisfying (\ref{experiment2}).
$\ncheck{-2}$ can be either 0 or 1. Suppose first that
$\ncheck{-2}=0$.  The first equation in (\ref{experiment5})
becomes $\dim \H{-1}_{\sss -}(\Qtilde) = \n{-1}-\n{0}$. There are only
two values for ${\bf\tilde n}$  for which
$\n{-1}-\n{0}\geq 0$ and these are $(1, 0, 0, 0, 1)$ and $(1, 1, 1, 0, 1)$:
for both of them $\dim \H{-1}_{\sss -}(\Qtilde)=0$. If
$\dim \H{-1}_{\sss -}(\Qtilde)=0$
the first two sequences in (\ref{experiment5}) split
into shorter ones and give
\be
\n{-1}=\ncheck{1}=0\qquad \n0 =\ncheck{-2}\qquad \n2=\ncheck0
\label{hminus10}
\ee
Therefore only 
\be
{\bf\tilde n}=(1, 0, 0, 0, 1)
\label{sol1}
\ee
is allowed, and for this solution we also have 
\be
{\bf\check{n}}\equiv (\ncheck{-2},\ncheck{-1},\ncheck{0},\ncheck1,\ncheck2)
=(0, 0, 1, 0, 1)
\label{sol1check}
\ee
thanks to (\ref{relation1}), (\ref{experiment4}) and (\ref{hminus10}).

Consider now the case $\ncheck{-2}=1$.  
The first equation in (\ref{experiment5})
becomes $\dim \H{-1}_{\sss -}(\Qtilde) = \n{-1}-\n{0}+1$. There are seven
values of ${\bf\tilde n}$ in the range (\ref{ranges}-\ref{inequalities}) 
for which the dimension of $\H{-1}_{\sss -}(\Qtilde)$
is non-negative: five of them  have $\dim \H{-1}_{\sss -}(\Qtilde)=0$ and two 
have $\dim \H{-1}_{\sss -}(\Qtilde)=1$.  Among the solutions with 
vanishing $\H{-1}_{\sss -}(\Qtilde)$ only two
are consistent with the relation (\ref{hminus10}) 
which derives from $\H{-1}_{\sss -}(\Qtilde)=0$. They are given by
\bea
&&{\bf\tilde n} = (1,0,1,0,0) \qquad {\rm with}\quad {\bf\check{n}}=(1,0,0,0,1)\label{sol2}\\
&&{\bf\tilde n} = (1,0,1,1,1) \qquad {\rm with}\quad {\bf\check{n}}=(1,1,1,0,1)
\label{sol3}
\eea
For the two values of ${\bf\tilde n}$ for which the dimension of 
$\H{-1}_{\sss -}(\Qtilde)$ is 1, the second equation in
(\ref{experiment5}) gives $\ncheck0=\ncheck1= 0,1$. Therefore 
to each of these two values of ${\bf\tilde n}$ there correspond two possible values for ${\bf\check{n}}$:
\bea
&&{\bf\tilde n} = (1,0,0,0,1) \qquad {\rm with}\quad {\bf\check{n}}=(1,0,0,0,1)
\label{sol4a}\\
&&{\bf\tilde n} = (1,0,0,0,1) \qquad {\rm with}\quad {\bf\check{n}}=(1,0,1,1,1)
\label{sol4b}\\
&&{\bf\tilde n} = (1,1,1,0,1) \qquad {\rm with}\quad {\bf\check{n}}=(1,0,0,0,1)
\label{sol5a}\\
&&{\bf\tilde n} = (1,1,1,0,1) \qquad {\rm with}\quad {\bf\check{n}}=(1,0,1,1,1)
\label{sol5b}
\eea

Finally, for each of the seven values of
${\bf\tilde n}$ and ${\bf \check{n}}$ listed in Eqs. (\ref{sol1}-\ref{sol5b}) 
we can compute the dimensions of
$\hhat{n}_{\sss 1,2}$ via the sequences (\ref{oursequences}). 
It turns out that 
Eq. (\ref{sol1}-{\ref{sol1check}),  Eq. (\ref{sol2})  and  Eq. 
(\ref{sol5a}) give rise to values for the dimensions of $\hhat{n}_{\sss 1,2}$
which are not consistent with the duality relations (\ref{poincare}).
Moreover, Eq. (\ref{sol3})
leads to $\dim \hhat{-1}_{\sss 1} = 1$. This implies that the dimension of
${\rm img}(\Wtilde{-2}; \Zcheck)$ is 1 
and thus $\dim\Wcheck{-2}=0$: but this is inconsistent with the fact that
$\ncheck{-2} =1$ in (\ref{sol3}).  

In conclusion no solution with $\H{-1}_{\sss -}(\Qtilde)=0$ is allowed.
There are only three acceptable values for ${\bf\tilde n}$ and 
${\bf \check{n}}$, those
listed in (\ref{sol4a}), (\ref{sol4b}) and (\ref{sol5b}): for all of them
\be
\dim \H{-1}_{\sss -}(\Qtilde) = \dim \H{-2}_{\sss -}(\Qtilde) =1\qquad{\rm
at }\quad p^2 =-{\bar m}^2 \approx -2.1 
\label{expsolution}
\ee
Among the three solutions, there is one, listed in (\ref{sol4a}), for
which $\htilde{n}=\htilde{-n}=\hcheck{n}$, for any $n$; thus this 
solution is consistent with $\Zcheck =0$, and with the possibility
that the $\Qtilde$ be related to a BRS operator with vanishing $\Zhat$
by a field redefinition which preserves Siegel gauge.  The other two
solutions have necessarily $\Zcheck$ and thus $\Zhat$ different than
zero.

\section{BRS Cohomologies without Gauge-Fixing}

In this Section we derive a relation between the cohomologies
$\H{n}(\Qtilde)$ at different ghost numbers by looking at the $p^2$
dependence of the operator $\Qtilde(p)$ acting on the 
non-gauge-fixed state spaces $\F{n}$. This relation is not implied by any of
the sequences that we constructed in the previous Section. The sequences of
the previous Section reflect properties of $\Qtilde$ at a {\it fixed} $p^2$.
The relation of this Section is instead
a consequence of the fact that the cohomology of $\Qtilde(p)$ is empty
for $p^2$ generic and it appears only on surfaces of positive
codimension in momentum space. The fact that this
relation is indeed satisfied by the solution (\ref{expsolution})
for the absolute cohomologies that we derived numerically
represents an independent check of the consistency of such a solution.

Let us fix in $\F{n}$ 
a basis $\{ v^{\sss (n)}_{i_n}(p)\}$ of vectors of momentum $p$.
Let $\Q{n}(p)$ be the matrix describing the action of
$\Qtilde$ on this basis
\be
\Qtilde(p)\, v^{\sss (n)}_{i_n}(p) =  \Q{n}_{i_{n+1}i_n}(p)\,  
v^{\sss (n+1)}_{i_{n+1}}(p)
\ee
The choice of the basis $\{ v^{\sss (n)}_{i_n}(p)\}$ is associated
with the choice of a {\it positive definite} hermitian product with
respect to which the basis is orthonormal. 
Let $\Qbar{n}(p)$ be the matrix which is the hermitian
conjugate of $\Q{n}(p)$: in the same basis,  $\Qbar{n}(p)$
represents the hermitian conjugate of 
$\Qtilde$  with respect to the positive definite hermitian product
defined above. Let us remark that the positive definite product
associated with the basis $\{ v^{\sss (n)}_{i_n}\}$
has nothing to do with the
bilinear form $(\cdot\,,\,\cdot)$ with respect to which $\Qtilde$ is
symmetric. The choice of the basis will play in this Section the
role that the choice of gauge had in Section 2.

Our discussion will focus on the hermitian matrix 
\be
\K{n}(p) \equiv \Qbar{n}(p)\,\Q{n}(p),
\ee 
Let ${\rm V}_{i_n j_n}$ be an invertible but not necessarily
unitary matrix. Under the change of basis 
\be
v^{\sss (n)}_{i_n}\to \sum _{j_n}\, {\rm V}_{i_n j_n}\, v^{\sss (n)}_{j_n},
\label{changeofbasis}
\ee
$\K{n}$ transforms as 
\be
\K{n}(p) \to  \overline{\rm V}\,\Qbar{n}(p)\, {\overline{\rm V}}^{-1}\, 
{\rm V}^{-1}\, \Q{n}(p)\, {\rm V}
\ee
This shows that, although eigenvalues and eigenstates of $\K{n}(p)$
are basis dependent, its kernel is not and it coincides with the
kernel of $\Qtilde$. 

Eigenstates of $\K{n}(p)$ with vanishing eigenvalues are of two types
(See Figure~\ref{f:typeAB.eps} (a)): (A) {\it generic} null eigenstates whose eigenvalues are zero for all
$p^2$; (B) eigenstates whose eigenvalues vanish for isolated values of
$p^2= -m^2$.  For reasons that we will review momentarily, eigenstates of $\K{n}(p)$ of type (B) are in the
cohomology of $\Q{n}$ at $p^2=-m^2$. Eigenstates of $\K{n}(p)$ of type
(A) are cohomologically trivial {\it except} for those values of $p^2$
for which $\K{n-1}(p)$ has eigenstates of type (B). To study the
cohomology of $\Qtilde$, it is therefore sufficient to determine the number 
$N^{\sss (n)}_B(p^2)$ of eigenstates of type (B) of ghost number $n$ 
whose eigenvalues vanish at a given value of $p^2$: 
the dimension of the cohomology of $\Qtilde(p)$ at ghost number $n$ is given by
\be
\dim \H{n}(\Qtilde(p)) = N^{\sss (n)}_B(p^2) + N^{\sss (n-1)}_B(p^2)
\label{continuity}
\ee
Note that the duality relation (\ref{absolutedual}) implies that
\be
N^{\sss (n)}_B(p^2) =N^{\sss (-n)}_B(p^2)
\ee
and thus knowledge of the  BRS cohomology at all ghost numbers only requires
the computation of $N^{\sss (n)}_B(p^2)$ for $n\leq 0$.
\begin{figure}
\begin{center}
\includegraphics*[scale=.42, clip=false]{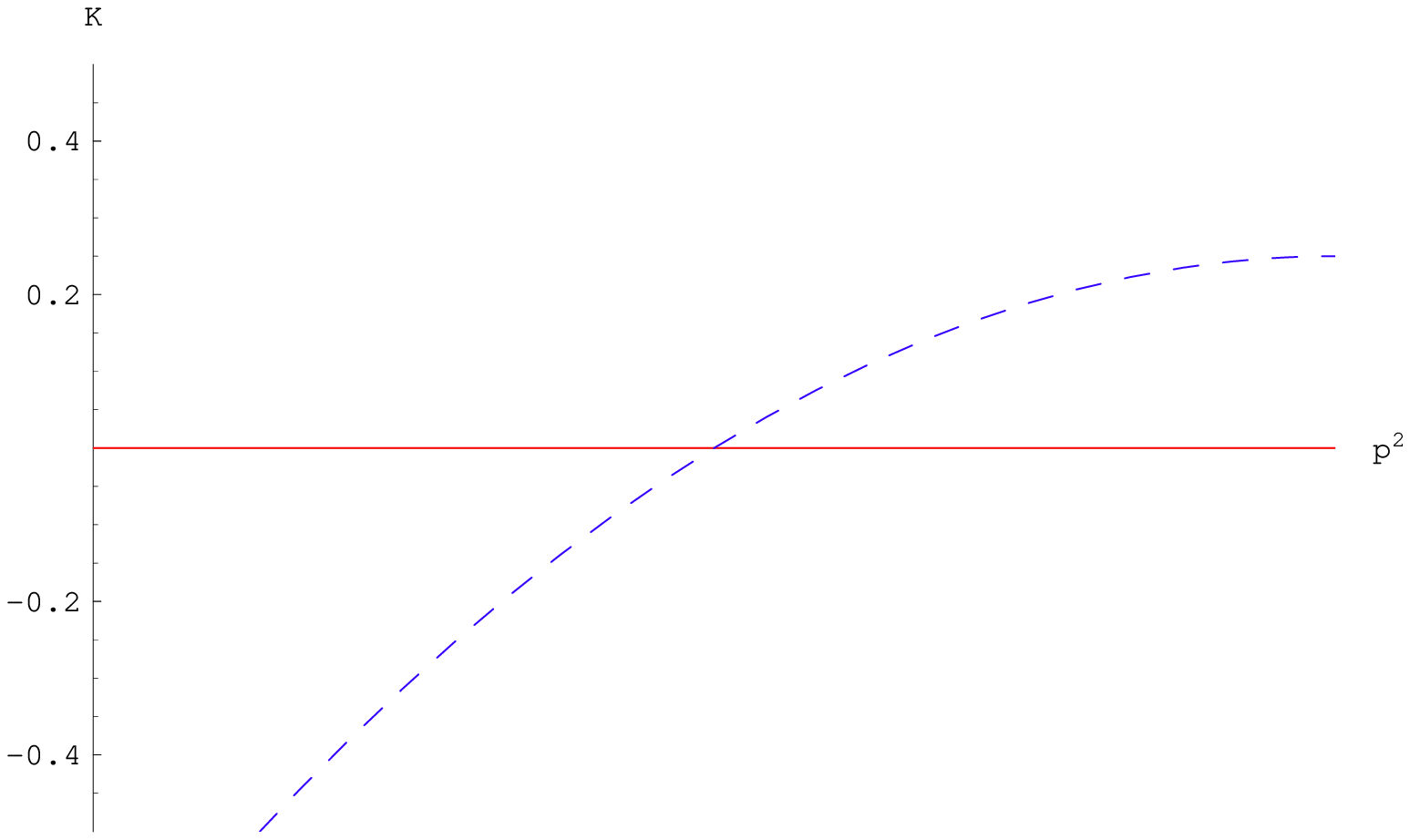} 
\includegraphics*[scale=.42, clip=false]{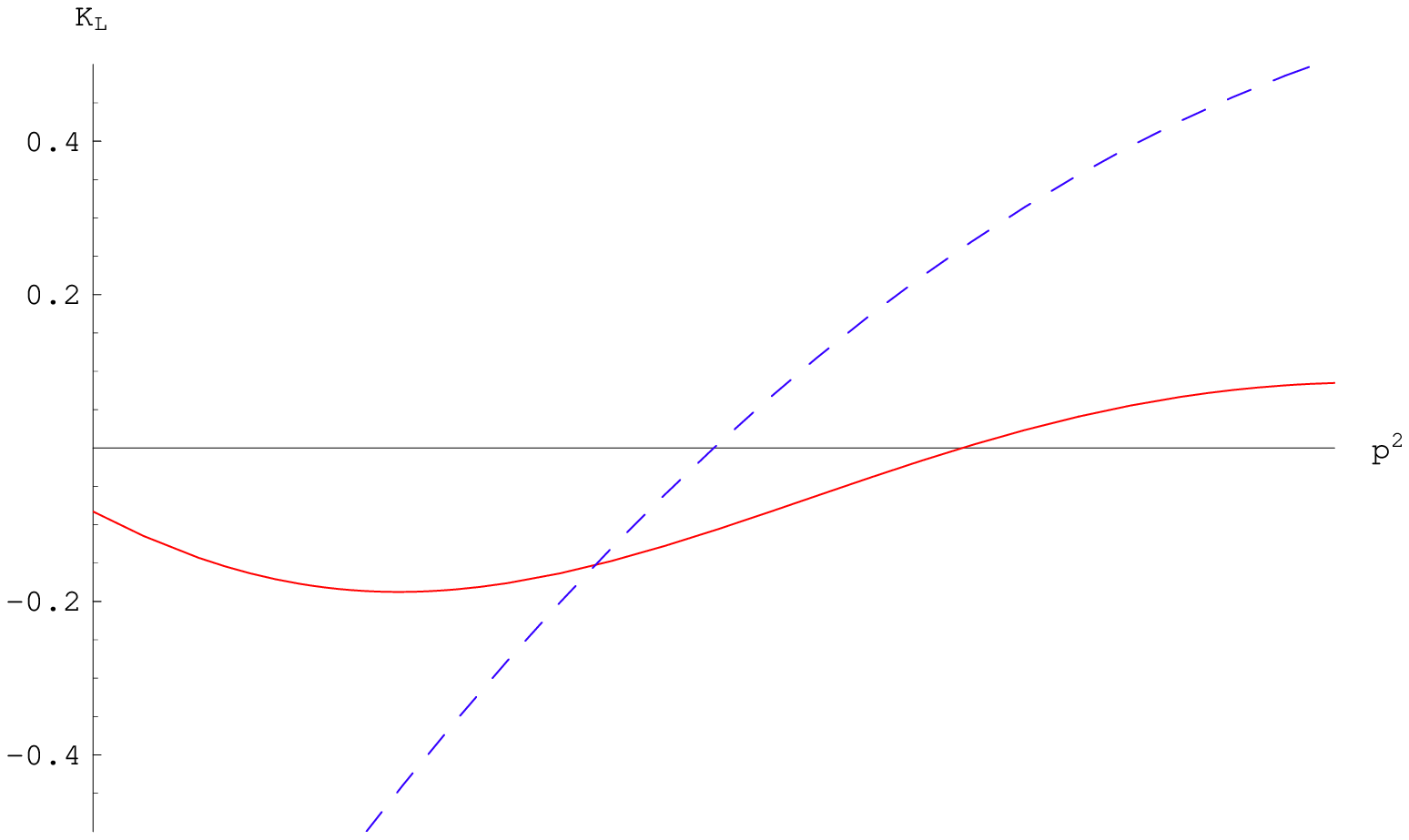}\\
{$\scriptstyle{\rm (a)}\hskip 6truecm\scriptstyle{\rm (b)}$}
\end{center}
\caption[x] {\footnotesize Eigenvalues of $\K{n}(p)$ (a) and of level truncated
$\K{n}_L$ (b) of type (A) (red) and of type (B) (dashed-blue).}
\label{f:typeAB.eps}
\end{figure}
Let us briefly review how (\ref{continuity}) derives from the continuity
of the spectrum of the hermitian and diagonalizable operator $\K{n}(p)$ 
as a function of $p^2$. Nilpotency of $\Qtilde$ means 
\be
\Q{n}(p)\, \Q{n-1}(p) =0,
\ee
i.e. the image of $\Q{n-1}(p)$ is contained in
the kernel of $\K{n}(p)$ for all $p^2$. For generic values of
$p^2$ the image of $\Q{n-1}(p)$ also coincides with the kernel of $\K{n}(p)$.
At some non-generic value of $p^2=-m^2$ two things can happen: either
a generically non-zero eigenvalue $\lambda(p)$ of $\K{n}(p)$ vanishes
at $p^2=-m^2$ (and the corresponding eigenstate is of type (B)); or
some generically trivial eigenstate (of type (A)) becomes non-trivial.  
In the first case, the type (B) eigenstate can be written as follows
\be
\psi_n(p) = {1\over\sqrt{\lambda(p)}}\, \Qbar{n} (p)\,\psi^\prime_{n+1}(p)
\ee
where
\be
\psi^\prime_{n+1}(p) \equiv {1\over\sqrt{\lambda(p)}}\,\Q{n}(p)\,\psi_n(p)
\ee
has unit norm for all $p^2$. This shows that $\psi_n(p)$ is
$\Qbar{n-1}$-closed for any $p^2$: moreover for $p^2=-m^2$,
$\psi_n(p)$ is also $\Q{n}$-closed. This implies that 
eigenstates of type (B) are
cohomologically non-trivial: indeed, a state that is both $\Qtilde$-trivial
and ${\overline\Qtilde}$-closed is orthogonal to itself with respect
with the positive hermitian product, and therefore it vanishes.
At the same time $\psi^\prime_{n+1}(p)$ is an eigenstate
of type (A) of $\K{n+1}(p)$: it lies generically in the image of
$\Qtilde$ and thus it is $\Qtilde$-closed for all $p^2$. Moreover at
$p^2=-m^2$ it is also ${\overline\Qtilde}$-closed and thus it is
cohomologically non-trivial at that value of $p^2$.  Summarizing,
eigenvalues of type (B) at ghost number $n$ which vanish at $p^2=-m^2$
are in one-to-one correspondence with eigenstates of type (A) at ghost
number $n+1$ which become non-trivial at the same value of $p^2$.

Suppose now that, according to Sen's hypothesis, $\H0(\Qtilde)=0$.
Then, at a given $p^2$, the relation (\ref{continuity}) implies
\be
0 = \dim \H{-1}(\Qtilde(p)) -\dim \H{-2}(\Qtilde(p)) +\cdots
\label{continuity2}
\ee
We saw that for $p^2= -{\bar m}^2\approx -2.1$, $\dim\H{-n}(\Qtilde)=0$
for $n\ge 3$. Therefore (\ref{continuity2}) predicts
\be
\dim \H{-1}(\Qtilde(-{\bar m}^2)) =\dim \H{-2}(\Qtilde(-{\bar m}^2))
\ee 
in agreement with (\ref{expsolution}).

\subsection{Numerical analysis}

The characterization of cohomologically non-trivial states
that we explained above leads to a method for the analysis of the 
BRS cohomology which is completely different than the one of Section 2. 
The method consists in calculating the number $N_B^{\sss (n)}$
of $\K{n}$ eigenstates of type (B) and using (\ref{continuity})
to evaluate the $\Qtilde$-cohomology. 

The problem, as usual, is that the level truncated BRS operator $\Q{n}_L(p)$ 
is only approximately nilpotent. Therefore the eigenvalues of type $(A)$ of
$\K{n}(p)$ become, in the LT approximation, generically non-vanishing,
and thus {\it a priori} indistinguishable from the eigenvalues of type
(B) (See Figure~\ref{f:typeAB.eps} (b)). To make use of (\ref{continuity})  one must find a way
to distinguish among the eigenvalues of $\K{n}_L$ those that 
correspond to eigenvalues of type (A) of the exact $\K{n}(p)$. 
With this aim, let us observe that for $p^2\gg 1$ the
non-perturbative $\Qtilde(p)$ converges exponentially to the
perturbative $Q(p)$. Since LT preserves the nilpotency of $Q(p)$, the
eigenvalues of type (A) of $\K{n}(p)$ correspond to eigenvalues of the
level truncated $\K{n}_L(p)$ which converge to zero for $p^2
\to\infty$. 

In conclusion the method should go as follows: one looks at an
eigenvalue $\lambda (p)$ of $\K{n}_L(p)$ that vanishes at some $p^2=-m^2$
and follows it for $p^2\gg-m^2$ into the
region where $\Qtilde(p)\approx Q(p)$. If the eigenvalue flows to zero
it is of type (A) and thus it does not contribute to $N_B^{\sss (n)}$; 
we will refer to such eigenvalues  as ``trivial''.
If on the other hand the eigenvalue diverges for $p^2\to\infty$ we will
call it ``non-trivial'': the corresponding eigenstate, for $p^2=-m^2$ 
is in the cohomology (of type (B)) of $\Qtilde$ at
ghost number $n$.

The method presents various technical difficulties. 
One difficulty is associated with ``level crossing'' of eigenvalues.
Suppose that at some $p_0^2> -m^2$ the eigenvalue
$\lambda_1(p)$ associated with the eigenstate $\psi_1(p)$ crosses another
eigenvalue $\lambda_2(p)$ corresponding to the eigenstate
$\psi_2(p)$. For $p^2> p_0^2$ one should be careful to follow the
eigenvalue corresponding to the eigenstate which is continuously
connected with $\psi_1(p)$ for $p^2<p_0^2$.
In the numerical situation authentic ``level crossing'' never
occurs. In the numerical approximation ``level crossing''  
appears as in Figure~\ref{f:levelcrossing.eps} (a): 
two eigenvalues $\lambda_1(p)$ and $\lambda_2(p)$
become almost degenerate
for $p^2 \approx p_0^2$ without ever coinciding;
the corresponding eigenstates $\psi_1(p)$ and $\psi_2(p)$ 
vary rapidly in the region $p^2\approx p_0^2$ and
switch among themselves when going through $p_0^2$:
\be
\psi_1(p)\bigr|_{p^2 =  p^2_0 -\epsilon} 
\approx \psi_2(p)\bigr|_{p^2 =  p^2_0 +\epsilon} 
\label{numericallc}
\ee
with $\epsilon$ positive and small. To characterize 
``numerical level crossing'' one necessitates a quantitative criterion
to decide what  ``rapid change'' of $\psi(p)$ means. For $p^2\approx p_0^2$
the two almost degenerate eigenstates $\psi_1(p)$ and $\psi_2(p)$
mix approximately only among themselves, and thus they can be written
as
\begin{figure}
\begin{center}
\includegraphics*[scale=.8, clip=false]{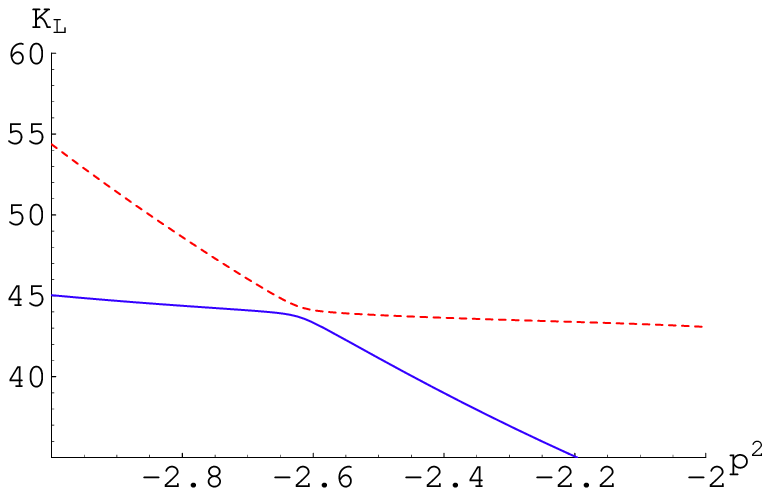}
\includegraphics*[scale=.8, clip=false]{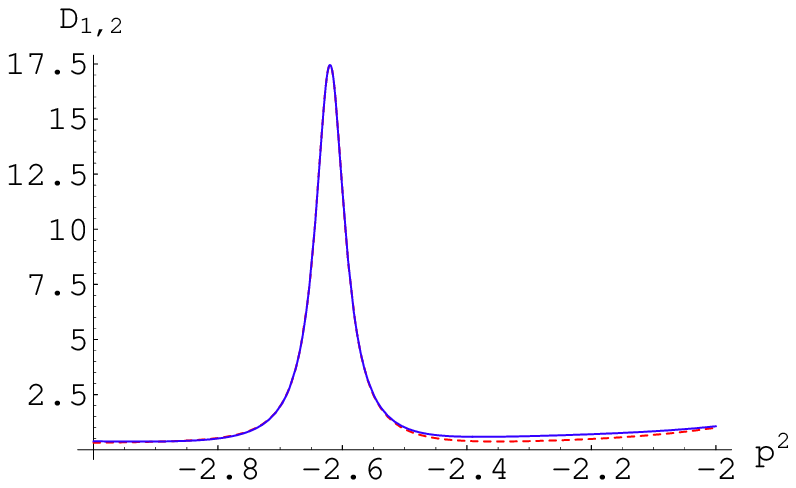}\\
{$\scriptstyle{\rm (a)}\hskip 6truecm\scriptstyle{\rm (b)}$}
\end{center}
\caption[x] {\footnotesize (a) Numerical ``level crossing'' of two 
eigenvalues of $\K0_L$ for $L=4$ in the even twist parity sector. (b) 
The functions $D_{1,2}(p^2)$ defined in Eq. (\ref{derivative}).}
\label{f:levelcrossing.eps}
\end{figure} 
\be
\psi_1(p) = \cos\theta(p)\, e_1 + \sin\theta(p)\, e_2
\qquad  \psi_2(p) = -\sin\theta(p)\, e_1 + \cos\theta(p)\, e_2
\ee
where $e_1 \equiv\psi_1(p^2_0-\epsilon)$ and 
$e_2 \equiv\psi_2(p^2_0 -\epsilon)$ are orthogonal. 
Hence the modulus of the derivatives of $\psi_1(p)$ and $\psi_2(p)$
\be
D_1(p^2) \equiv \Bigl\Vert{d\, \psi_1(p)\over d\, p^2}\Bigr\Vert^2\qquad
D_2 (p^2) \equiv\Bigl\Vert{d\, \psi_2(p)\over d\, p^2}\Bigr\Vert^2
\label{derivative}
\ee
are approximately coincident in the region $p^2\approx p_0^2$
\be
D_1(p^2)\approx D_2(p^2) \approx \Bigl({d\,\theta(p)\over d\,p^2}\Bigr)^2 
\ee
Sharp peaks of the function above can be taken as the signals of ``numerical
level crossing'' (Figure~\ref{f:levelcrossing.eps} (b)). 
It is clear that this definition of level
crossing involves some arbitrariness: peaks of the function (\ref{derivative})
can be more or less sharp corresponding to more or less exact
exchange of eigenstates when going through the almost-degeneracy region.
\begin{figure}
\begin{center}
\includegraphics*[scale=.8, clip=false]{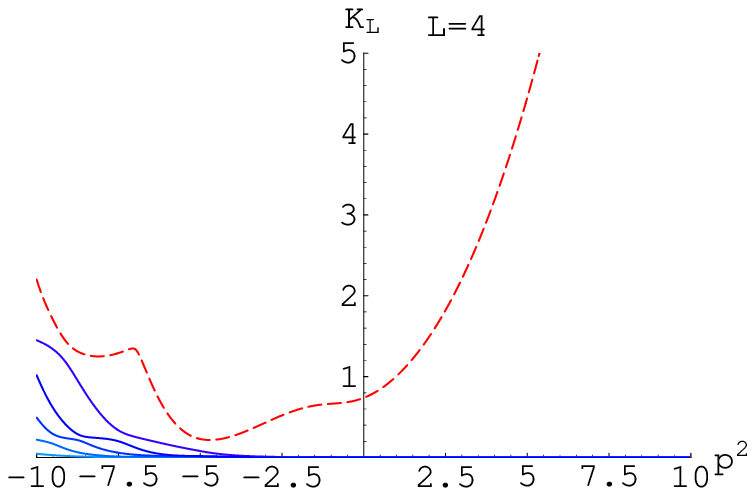}
\includegraphics*[scale=.8, clip=false]{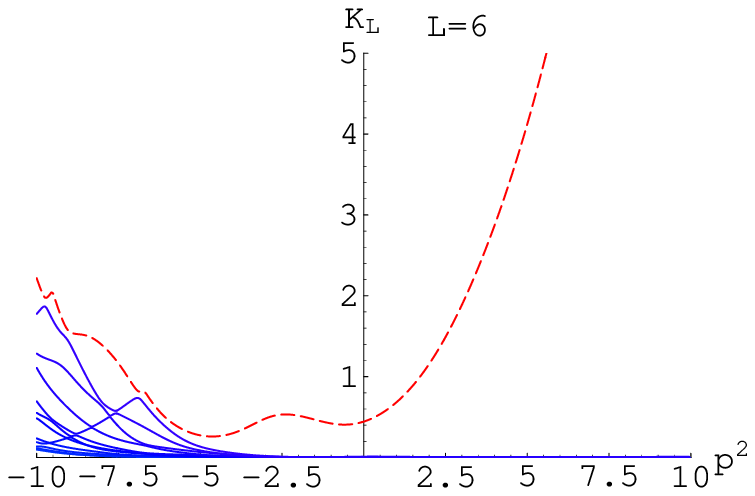}
\end{center}
\caption[x] {\footnotesize Lowest non-trivial eigenvalue (dashed-red) and
trivial eigenvalues 
of $\K{0}_L(p)$ in the even parity sector for $L=4,6$.}
\label{f:notgaugefixed0even.eps}
\end{figure}
\begin{figure}
\begin{center}
\includegraphics*[scale=.8, clip=false]{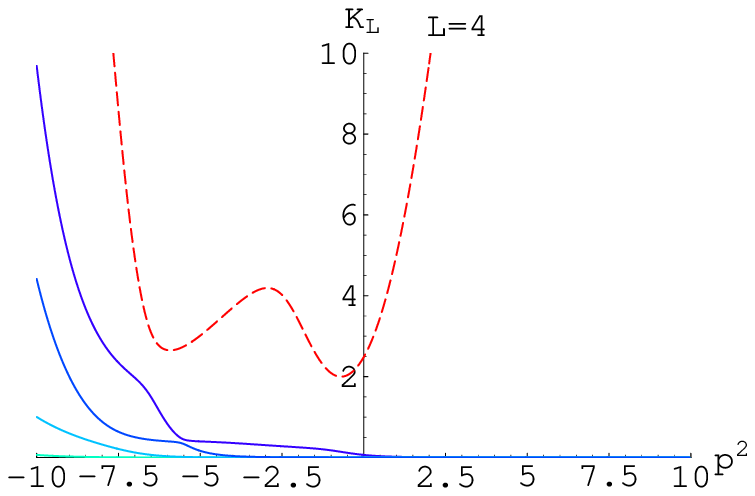}
\includegraphics*[scale=.8, clip=false]{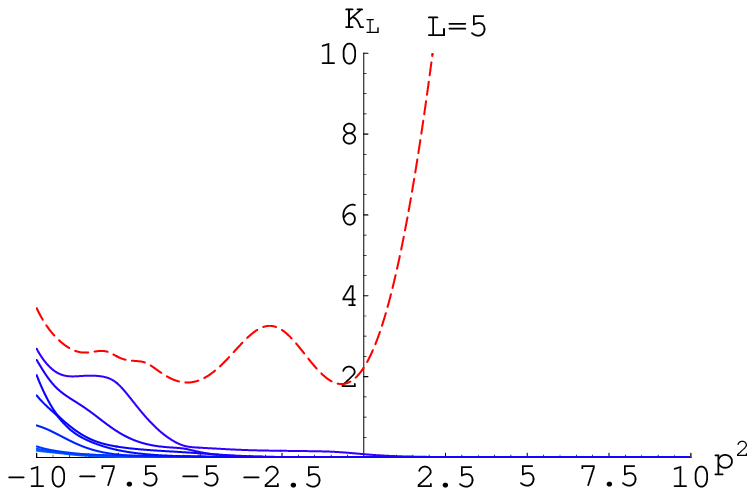}
\includegraphics*[scale=.8, clip=false]{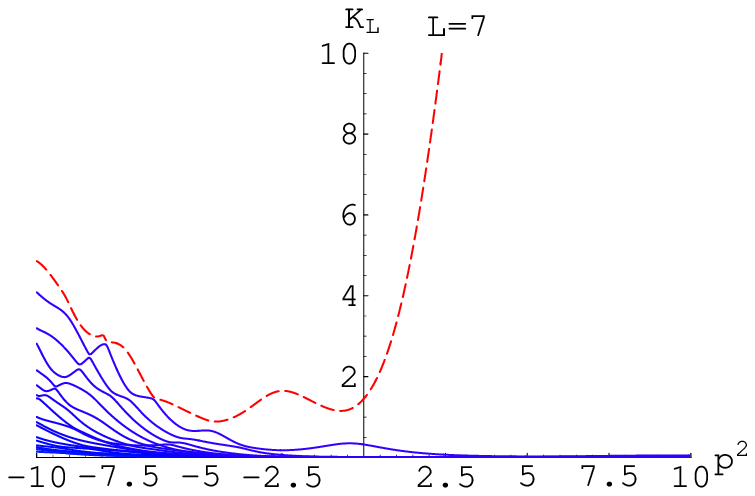}
\end{center}
\caption[x] {\footnotesize Lowest non-trivial eigenvalue (dashed-red) and
trivial eigenvalues 
of $\K{0}_L(p)$ in the odd parity sector for $L=4,5,7$.}
\label{f:notgaugefixed0odd.eps}
\end{figure}
Another practical disadvantage of this method is the following: when
the level increases BRS nilpotency is more accurate for a wider range
of negative $p^2$. Hence the zeros of the  eigenvalues that are approximately
of type (A) become more and more dense on the real (negative) axis. For big
enough levels there is not only an ever increasing number of eigenvalues
vanishing at some $p^2$ to be followed into the perturbative region: 
also the number of level crossings for each eigenvalue
grows rapidly, making progressively more cumbersome to determine
if the vanishing eigenvalues are trivial or not.
In practice one can more easily determine a region of the $p^2$-axis
where the non-trivial eigenvalues 
do not vanish and the numbers $N_B^{\sss (n)}$ are zero. 

This method is also not well suited to compute $N_B^{\sss (n)}$ for $n\not=0$. 
This is so because the level truncated matrix $\Q{n}_L(p)$ is a square
matrix only for $n=0$: for $n < 0$ the number of rows is bigger than
the number of columns since the number of states of level $L$ and
ghost number $n$ is less than the number of states of level $L$ 
and ghost number $n+1$. For a generic $m\times n$ matrix depending
on the parameter $p^\mu$ the condition for non-empty kernel determines
a sub-manifold of codimension greater than 1 in $p$-space, if $m>n$.
Therefore, in the numerical approximation, eigenvalues of $\K{n}_L(p)$
never exactly vanish on the $p^2$ axis when $n\not=0$: 
the number $N^{\sss (n)}_B$
should be rather identified with the number of eigenvalues which diverge
for $p^2\to\infty$ and are ``almost'' vanishing for some real $p^2$
\footnote{The method for computing the BRS cohomology that we
describe in this Section is somewhat similar in spirit to the one in
\cite{et}. Both methods look at the kernel of $\Q{n}_L(p)$ and
try to determine, in different ways, which of its elements are
``approximately'' trivial. Both methods however are able to evaluate
cohomology of type (B) only: eigenstates of type (A) that become
non-trivial for some isolated values of $p^2$ correspond, in the
numerical approximation, to eigenvalues that are generically
non-vanishing. If one looks at them when they vanish, they might well
be ``approximately'' trivial even if they are not so at some other
value of $p^2$ where they do not vanish. In other words cohomology
associated to states of type (A) that become non-trivial for isolated
values of $p^2$ is invisible within methods that look at the kernel of
$\Q{n}_L(p)$. If one were able to compute type (B) cohomology at all
ghost numbers, this would not be a limitation, thanks to
(\ref{continuity}). But we just explained that for $n\not=0$ the
numerical methods that study the kernel of $\Q{n}_L(p)$ become only
qualitatively meaningful.}.

In conclusion, the practical relevance of  this method is limited 
to determining the region on the $p^2$ axis for which it is safe to say that
$N^{\sss (0)}_B$ is zero. 

Let us describe the results we obtained. We studied the spectrum of
$\K{0}_L(p)$ for levels $L=4,5,6,7$, both in the even and the odd twist
parity sectors. The numbers of states for each level 
are reported in Table II. The results of our computations
are shown in 
Figures~\ref{f:notgaugefixed0even.eps} and \ref{f:notgaugefixed0odd.eps}. 
The perturbative region
for which $\Qtilde(p)\approx Q(p)$ is for $p^2>10$. 
The ``non-trivial'' eigenvalues remain separate from type (A) up
to $p^2\approx -6$. This excludes cohomology of type (B) of ghost number
0 for $p^2 > -5$, in agreement with the results of Section 3. 
The results for ghost numbers -1 and -2 are less 
transparent. One has also to take into account that 
there are much less states at these ghost numbers than at 
ghost number 0: so we expect the
LT approximation to be less accurate. The analysis of Section 4 
gave $N^{\sss (-1)}_B=0$ and $N^{\sss (-2)}_B=1$ for $p^2\approx -2$ 
in the odd sector. Although we find that there is a ``non-trivial''
eigenvalue that has a minimum at $p^2\approx -2$ it does not seem that this
minimum becomes more pronounced as the level is increased
(Figure~\ref{f:notgaugefixedminus2.eps}).
$$
\vbox{\tabskip=0pt
\setbox\strutbox=\hbox{\vrule height12pt depth8pt width0pt}
\halign{\strut#& \vrule#& \hfil#\hfil &
\vrule#& \hfil#\hfil&\vrule#& \hfil#\hfil & \vrule#& \hfil#\hfil & \vrule#& \hfil#\hfil & \vrule#\tabskip=0pt\cr
\noalign{\hrule}
& & Level & & ~ghost \# 0~ & &~ghost \# -1~ & &~ghost \# -2~ & &~ghost \# -3
~&\cr\noalign{\hrule}
& & ~3 (odd)  & & 15   & & 7 & & 1 & & 0 &\cr\noalign{\hrule}
& & ~4 (even) & & 37  & & 15 & & 2 & & 0 &\cr\noalign{\hrule}
& & ~5 (odd) & & 75  & & 37 & & 7 & & 0 &\cr\noalign{\hrule}
& & ~6 (even) & & 150  & & 75 & & 15 & & 1 &\cr\noalign{\hrule}
& & ~7 (odd)  & & 308 & & 160 & & 37 & & 2 &\cr\noalign{\hrule}
}}
$$
\nobreak
\centerline{Table II: Number of scalar states at various levels.}
\bigskip
\begin{figure}
\begin{center}
\includegraphics*[scale=.8, clip=false]{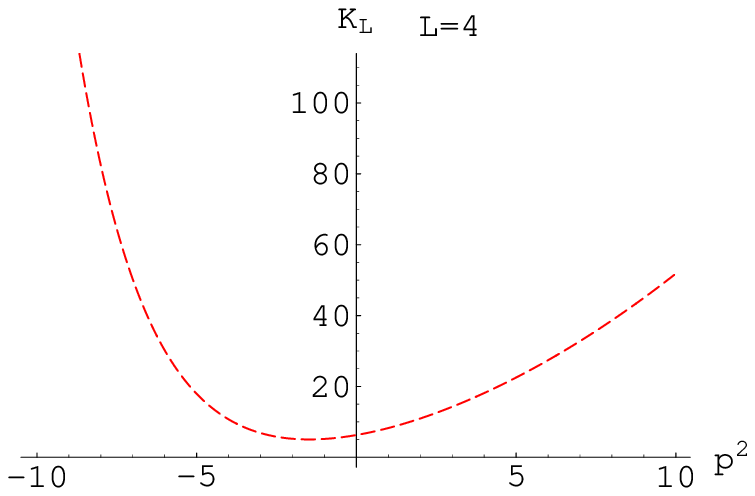}
\includegraphics*[scale=.8, clip=false]{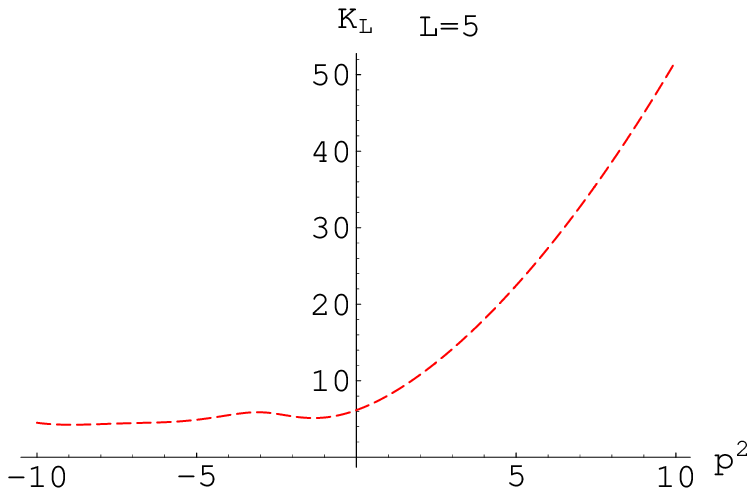}
\includegraphics*[scale=.8, clip=false]{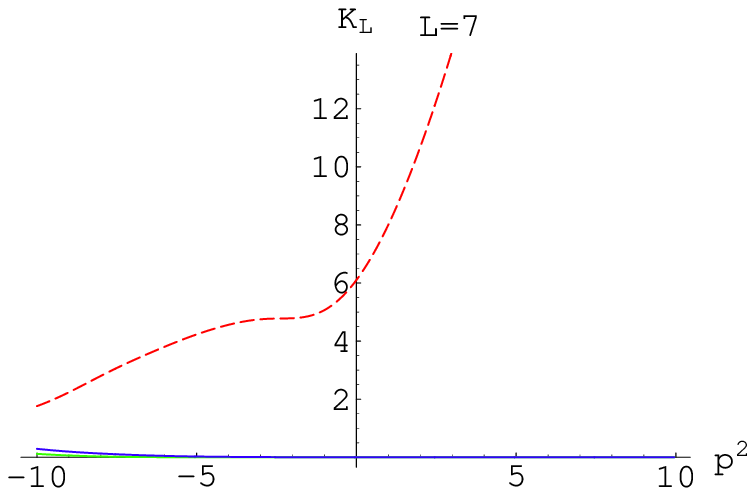}
\end{center}
\caption[x] {\footnotesize Lowest non-trivial eigenvalue (dashed-red) and
trivial eigenvalues 
of $\K{-2}_L(p)$ in the odd parity sector for $L=4,5,7$.}
\label{f:notgaugefixedminus2.eps}
\end{figure}
\section{Conclusions}
In this paper we presented a method for the computation of the number of 
physical states in OSFT quantized around the tachyonic vacuum, within LT
approximation scheme.

We explained why any attempt to compute the BRS cohomology by looking at the 
kernel of the (approximately nilpotent) level truncated BRS operator $\Qtilde$
is plagued by some intrinsic limitations. By such methods one can only 
compute a certain subset --- denoted as ``of type B'' in the text --- of the
ghost number 0 cohomology; moreover these methods become more and more 
inefficient as the level increases.

We thus developed a computational scheme that appears to be better suited to 
LT approximation. The method focuses on the kinetic operators, 
$\Ltilde^\ss{n}$, of the {\it gauge-fixed} OSFT action expanded around 
the tachyonic vacuum, both in the matter ($n=0$) and in the various ghost 
($n\not=0$) sectors. In contrast to the kernel of $\Qtilde$, the kernels of 
$\Ltilde^\ss{n}$ are generically empty, as a consequence of gauge-fixing,
and acquire a non-vanishing dimension at isolated values of the space-time 
squared momentum $p^2$. For this reason, zeros of $\det \Ltilde^\ss{n}$ in the level
truncated theory are expected to be stable as the level varies, for those
values of $p^2$ ($p^2\gg -2L$) where LT approximation should apply. 
We performed a numerical computation of $\Ltilde^\ss{n}$ up to level 9, in
Siegel gauge and in the scalar sector of the theory: this computation
confirmed the expectation above, even if the range of $p^2$ where LT 
approximation seems to be accurate is somewhat smaller than expected:
$p^2\gtrsim -5$ for $L=9$.

We used the numerical data concerning the vanishing spectrum of
$\Ltilde^\ss{n}$ in two ways. To begin with, we expressed, by means of
the Fadeev-Popov formula, the dimension of the physical state space of
OSFT as an index, constructed out of the numbers of zeros of $\det
\Ltilde^\ss{n}(p)$ weighted with their multiplicities.  In the region
of $p^2$ for which LT approximation is valid, there is a single group
of zeros of the determinants $\det \Ltilde^\ss{n}(p)$ centered around
$p^2={\bar m}^2 \approx -2.1$, whose spread decreases as the level
goes up, and whose Fadeev-Popov index vanishes. It is reasonable
to conclude that this group of zeros corresponds in the exact
theory to a multiplet of degenerate matter and ghost fields 
carrying no physical degree of freedom. This is our numerical
evidence confirming Sen's conjecture that there are no open string
states around the tachyonic vacuum.

Assuming that the group of zeros of $\det \Ltilde^\ss{n}(p)$ at
$p^2={\bar m}^2 \approx -2.1$ really corresponds to an exactly
degenerate multiplet, we were also able to prove that, at the same
$p^2$, some of the negative ghost number BRS cohomologies are
non-empty: $\dim \H{-1}(\Qtilde(-{\bar m}^2)) =\dim
\H{-2}(\Qtilde(-{\bar m}^2))=1$.

This result derives from two circumstances: first, the dimensions of
the kernels of the kinetic operators $\Ltilde^\ss{n}$ are connected
with the dimensions of the $\Qtilde$ cohomologies {\it relative} to
$b_0$; second, the relative $\Qtilde$-cohomologies are related to the
absolute $\Qtilde$-cohomologies $\H{n}({\Qtilde})$, although we
emphasized that this relation, in the non-perturbative case, is
considerably more involved than in the perturbative one. In this paper
we derived the non-perturbative long exact sequence
(Eq. (\ref{nonpertbottsequence})) which connects absolute and relative
BRS cohomologies, together with two ``sister'' long exact sequences
involving some new kind of relative BRS cohomologies
(Eq. (\ref{oursequences})): this is an exact result, independent of the
LT approximation.

Let us mention some possible extensions of our work. From a technical point
of view, it would be, of course, very useful to improve the accuracy and
to extend the $p^2$-range of validity of our approximation, both by using
more powerful and efficient computational tools and by means of extrapolation
algorithms like the ones in \cite{gr} and \cite{beccaria}.
One should also consider the extension of our computation to the states
of higher space-time spin, in particular with the purpose of investigating 
the gauge field sector of the string theory around the tachyonic vacuum. The
main problems to face in order to carry out this program are again of mere 
computational type.

From a more conceptual point of view, the obvious question that our results 
raise is the physical meaning of the BRS cohomology at negative ghost numbers.
The fact that such cohomology is non-empty is a novel feature of the 
non-perturbative theory, with respect to the perturbative one; even if it does
not contradict the original conjecture of Sen which identifies 
the tachyonic vacuum with the closed string vacuum, it does not agree with the 
stronger hypothesis of Vacuum SFT \cite{rsz}, 
according to which the BRS operator
around the tachyonic vacuum has empty cohomology at all ghost numbers.

\section*{Acknowledgments} 

We thank C. Becchi, L. Rastelli and A. Sen 
for useful conversations. We are grateful to L. Bertora for writing a code
that helped us to improve our numerical results, to M. Beccaria for
explaining us some issues regarding Mathematica programming, 
to I. Ellwood and W. Taylor for sending us some of their numerical data 
and to S. Pinsky for putting at our disposal his computing facilities.
We would also like to thank the organizers of the Workshop on
String Theory at the Harish-Chandra Research Institute, Allahabad
for allowing us to present and discuss part of the results of this paper in a
stimulating scientific environment.
This work is supported in part by
Ministero dell'Universit\`a e della Ricerca Scientifica e Tecnologica
and the European Commission's Human Potential program under contract
HPRN-CT-2000-00131 Quantum Space-Time, to which the authors are
associated through the Frascati National Laboratory.

\end{document}